\documentclass[twocolumn]{aastex631}

\usepackage{multirow}
\usepackage{here}
\usepackage{graphicx}
\usepackage{tablefootnote}
\usepackage{tabularray}
\usepackage{url}
\usepackage{hyperref}

\begin{document}

\title{Predicting Dust Temperature from Molecular Line Data
\\Using Machine Learning}

\correspondingauthor{Tenta Dougome} 
\email{k9464116@kadai.jp} 
\author{Tenta Dougome} 
\affiliation{Department of Physics and Astronomy, Graduate School of Science and 
Engineering, Kagoshima University, 1-21-35 Korimoto, Kagoshima, Kagoshima 890-0065, 
Japan} 
\author{Yoshito Shimajiri} 
\affiliation{Kyushu Kyoritsu University, Jiyugaoka 1-8, Yahatanishi-ku, Kitakyushu, Fukuoka, 
807-08585, Japan} 
\author{Kazuya Saigo} 
\affiliation{Department of Physics and Astronomy, Graduate School of Science and 
Engineering, Kagoshima University, 1-21-35 Korimoto, Kagoshima, Kagoshima 890-0065, 
Japan} 
\author{Sanemichi Takahashi} 
\affiliation{Department of Physics and Astronomy, Graduate School of Science and 
Engineering, Kagoshima University, 1-21-35 Korimoto, Kagoshima, Kagoshima 890-0065, 
Japan} 
\author{Miyu Kido} 
\affiliation{Department of Physics and Astronomy, Graduate School of Science and 
Engineering, Kagoshima University, 1-21-35 Korimoto, Kagoshima, Kagoshima 890-0065, 
Japan} 
\author{Shu Ishibashi} 
\affiliation{Department of Physics and Astronomy, Graduate School of Science and 
Engineering, Kagoshima University, 1-21-35 Korimoto, Kagoshima, Kagoshima 890-0065, 
Japan} 
\author{Shigehisa Takakuwa} 
\affiliation{Department of Physics and Astronomy, Graduate School of Science and 
Engineering, Kagoshima University, 1-21-35 Korimoto, Kagoshima, Kagoshima 890-0065, 
Japan}  
\affiliation{Academia Sinica Institute of Astronomy \& Astrophysics, 11F of Astronomy
Mathematics Building, AS/NTU, No.1, Sec. 4, Roosevelt Road, Taipei 10617, Taiwan, ROC}

\begin{abstract}
We conducted experiments with machine learning techniques to construct dust temperature maps from the CO isotopologue molecular line data in the Orion A molecular cloud. In the classical astrophysical methodology, multi-band continuum data are required to derive the dust temperature. The present study aims to investigate the capability and limitations of machine learning techniques to derive dust temperatures in regions without multi-band dust continuum data.
We investigated how the number of pixels used for training influences prediction accuracy, and how the dust temperatures sampled in the training area influence prediction accuracy. We found that $\sim$5\% of the total number of pixels in the observational region is sufficient for training to obtain accurate predictions.
Furthermore, a dust temperature sample within the training area should cover the whole temperature range and have a similar sample distribution to that of the entire observing region for an accurate prediction.
The $^{12}$CO / $^{13}$CO ratio is often found to be the most important feature in predicting the dust temperature. As the $^{12}$CO / $^{13}$CO ratio is a tracer of PDR, the machine learning technique could connect the dust temperatures to the PDRs.
We also found that the condition of thermal gas-dust coupling is not required for accurate prediction of the dust temperature from the molecular line data, and that machine learning is capable of capturing information more than classical astrophysical concepts.
\end{abstract}


\section{Introduction} \label{sec:intro}

Dust temperatures in star-forming regions are used to measure the mass of molecular clouds and to understand their physical condition. Dust temperatures can be estimated from spectral energy distributions (SEDs) fitting to multi-wavelength continuum data in the far-infrared and submillimeter bands \citep{Roy13_Herschel}. The Herschel Gould Belt Survey (HGBS) conducted multi-wavelength observations at 70, 100, 160, 250, 350, and 500 $\mu$m toward various star-forming regions, including Orion A, B, Taurus, Ophiuchus, Lupus, Pipe Nebula, Aquila, Musca, and Perseus, and estimated the dust temperature in each region \citep{Roy13_Herschel, Schneider2013, Kirk2013, Rygl2013, Peretto2012, Konyves2015, Cox2016, Pezzuto2012}.
However, the estimate of dust temperatures toward the regions where HGBS did not observe is limited. 

The Herschel space infrared telescope is no longer in operation.
Although it is possible to obtain continuum data using ground-based telescope cameras, the process of removing atmospheric signals results in the loss of information on the extended structure \citep{Enoch2006, Shimajiri11, Shimajiri2015, Mattern2024}.
This occurs because signals from large-scale spatial structures appear as components common to many detectors. In the data reduction process, these common-mode components are removed as atmospheric signals, resulting in the loss of the extended real astronomical emission.
Therefore, estimating a dust temperature on a molecular cloud scale only with data obtained from ground-based single-dish telescope cameras is not desirable.
In contrast, ground-based radio telescopes can obtain molecular line data, providing the data with high angular resolution and extended structures.

Applications of machine learning techniques to observational data have been attractive in modern astronomy.
For example, \citet{Fujita2023} adopted a machine learning technique to solve the degeneracy of the kinematical distance to molecular clouds in the inner Galaxy.
\citet{Barchi2020} improved the accuracy of galaxy morphological classification using deep learning.
\citet{Schanche2019} combined random forests and deep learning methods to classify whether an object is a transiting exoplanet or not.
\citet{Zavagno2023} developed a supervised learning model to identify filamentary structures within galaxies.
\citet{Ueda2020} and \citet{Nishimoto2025} developed a model to identify infrared ring structures with a Convolutional Neural Network (CNN) and a Single Shot MultiBox Detector (SSD), respectively. The SSD model reduces processing time while maintaining accuracy compared to a CNN model. 
\citet{Gratier21} and \citet{shimajiri23} successfully predicted H$_{2}$ column densities from molecular line data. These results prove that the machine learning technique is now a powerful tool for astrophysics.

In this study, we aim to evaluate the effectiveness and limitations of machine learning techniques in predicting dust temperatures from molecular line data, focusing on the Orion A region as a case study.
We will make a model using Extra Trees Regressor \citep{Geurts2006}, which is similar to the Random Forest, by using molecular line data and dust temperatures as training data. This model learns the relationship between values in molecular line data and dust temperatures and enables us to predict dust temperatures from the molecular line data.

The structure of this paper is as follows. Section \ref{sec: Methods and Experiments} provides a detailed description of the data used for machine learning and the techniques employed to construct the machine learning model. The results of the dust temperature predictions are presented in Section \ref{sec: results}.
Section \ref{sec: discussion} investigates the dependence of the prediction accuracy on various factors. Specifically, we examine the proportion of data that must be designated as the training set to achieve reliable accuracy. We also compare the performance of the models
with and without the photon-dominated regions (PDRs).
Correlation of the prediction accuracy with the visual
extinction $A_{V}$ is also investigated.
Section \ref{sec: summary} summarizes this study.

\section{Methods and Experiments} \label{sec: Methods and Experiments}
\subsection{Data set} \label{sec: data set}
\begin{figure*}[htbp]
    \centering
    \includegraphics[width=17cm]{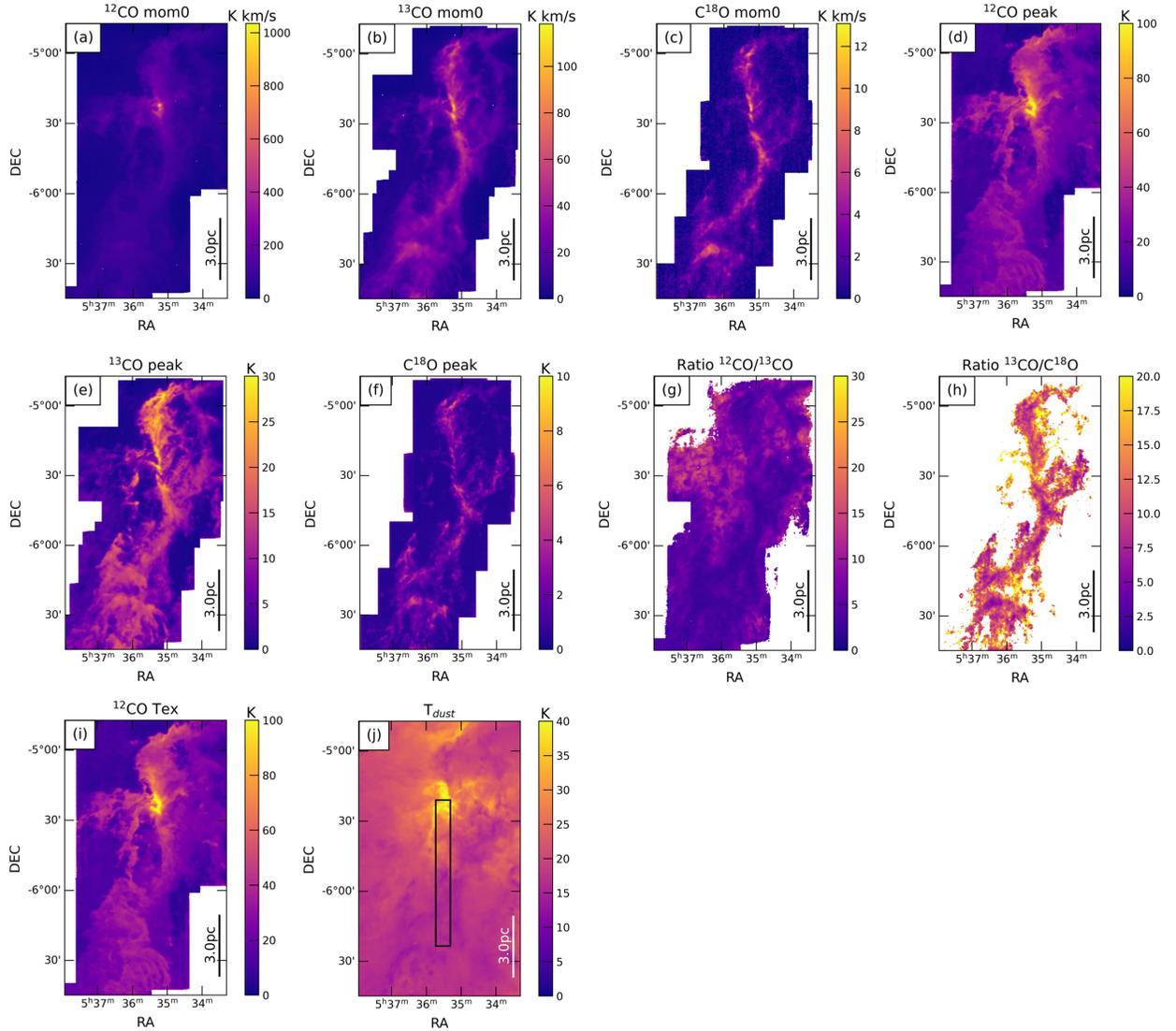}
    \caption{(a--c) $^{12}$CO (1--0), $^{13}$CO (1--0), and C$^{18}$O (1--0) integrated intensity (moment 0) maps, (d--f) $^{12}$CO, $^{13}$CO, and C$^{18}$O peak intensity maps, (g--h) maps of the integrated intensity ratio of $^{12}$CO to $^{13}$CO and of $^{13}$CO to C$^{18}$O, (i) excitation temperature map as derived from the $^{12}$CO peak, and (j) HGBS dust temperature map toward the Orion A molecular cloud. The angular resolution of all maps is 36$\arcsec$. Black box indicates the training area for Regressor-Region-A.}
    \label{fig0}
\end{figure*}

For our machine learning experiment, the archival mapping data of $^{12}$CO (1--0), $^{13}$CO (1--0), and C$^{18}$O (1--0) toward the Orion A molecular cloud taken with the Nobeyama 45-m telescope at an angular resolution of 21$\arcsec$.7 are adopted \citep[see Fig. \ref{fig0}, ][]{Shimajiri11, Shimajiri14_13CO/C18O, Nakamura19}. The dust temperature map in the same region as that of the molecular line data is produced from the data (70, 100, 160, 250, 350, and 500 $\mu$m) obtained from the Herschel space infrared telescope \citep{Andre10}.
The molecular-line maps are gridded to have the same pixel size of 7.5$\arcsec$,
and convolved with a Gaussian beam to have the same beam size of 36$\arcsec$
as that of the Herschel images.

\subsection{Machine Learning} \label{sec: Machine learning}
Machine learning can be categorized into three main types: supervised learning, unsupervised learning, and reinforcement learning.
This study adopts supervised learning, which uses two types of data: features (input variables) and target values (observed true values).
By providing a supervised learning algorithm with these data, a model is trained to obtain the relationship between the features and the target values.
In our specific study, the feature data correspond to various molecular-line images made from the 45-m CO isotopologue mapping data in Orion A, while the target values correspond to the dust temperature map made from the multi-wavelength Herschel images.

A small sub-region in the observational maps
(typically $\sim$5\% of the entire pixels) is defined as a training area.
Within that training area, pixel-by-pixel
relationships between the dust temperatures
and the selected molecular-line features are examined, and the models
that connect the dust temperatures and the molecular-line features
are constructed. The models are then applied to 
the molecular-line data across the entire map to
predict the dust temperature in each pixel.

\begin{table*}
    \centering
    \caption{Example of the pixel values of the target and feature data used for the training in Regressor-Region-A}
    \label{tab:example_training_table}
    \scriptsize
    \begin{tabular}{cccccccccccc} 
        \hline
        \multirow{2}{*}{RA$_{\mathrm{pix}}$} &\multirow{2}{*}{DEC$_{\mathrm{pix}}$} & $^{12}$CO & $^{13}$CO & C$^{18}$O & $^{12}$CO & $^{13}$CO & C$^{18}$O &\multirow{2}{*}{$ \frac{\rm ^{12}CO~mom0}{\rm ^{13}CO~mom0}$}&\multirow{2}{*}{$\frac{\rm ^{13}CO~mom0}{\rm C^{18}O~mom0} $} &\multirow{2}{*}{$T_{\rm ex}$} &\multirow{2}{*}{$T_{\rm dust}$} \\
        &&mom0&mom0&mom0&peak&peak&peak& \\
        & & [K km s$^{-1}$] & [K km s$^{-1}$] & [K km s$^{-1}$] & [K] & [K] & [K] &&&[K]& [K]  \\
        \hline
    400 & 400 & 80.067 & 24.339 & 2.0832 & 26.697 & 12.855 & 1.760 & 3.290 & 11.683 & 30.218 & 16.182 \\
    400 & 401 & 80.263 & 24.325 & 2.1518 & 26.495 & 12.506 & 1.793 & 3.300 & 11.304 & 30.015 & 16.226 \\
    400 & 402 & 80.850 & 24.331 & 2.1139 & 26.461 & 12.333 & 1.802 & 3.323 & 11.510 & 29.981 & 16.255 \\
    400 & 403 & 81.671 & 24.283 & 2.0156 & 26.796 & 12.235 & 1.776 & 3.363 & 12.047 & 30.317 & 16.286 \\
    400 & 404 & 82.642 & 24.203 & 1.9620 & 27.254 & 12.250 & 1.717 & 3.415 & 12.335 & 30.776 & 16.324 \\
    ... & ... & ... & ... & ... & ... & ... & ... & ... & ... & ... & ... \\
        \hline
    \end{tabular}
    \par
    \raggedright From left to right: the coordinates in pixel, the integrated intensity and peak intensity of $^{12}$CO, $^{13}$CO, and C$^{18}$O, the ratio of $^{12}$CO to $^{13}$CO and $^{13}$CO to C$^{18}$O, the excitation temperature, and the HGBS dust temperature.
\end{table*}

\subsection{Data Preprocessing} \label{subsec: Data pre}
As feature data, nine different types of molecular line maps are generated from the original CO-isotopologue image cubes. Those include $^{12}$CO, $^{13}$CO, and C$^{18}$O integrated intensity (mom 0) maps, $^{12}$CO, $^{13}$CO, and C$^{18}$O peak intensity maps, maps of the excitation temperature ($T_{\rm ex}$) derived from $^{12}$CO, and integrated intensity ratios of $^{12}$CO / $^{13}$CO and $^{13}$CO / C$^{18}$O (see Fig. \ref{fig0}, Table \ref{tab:example_training_table}). The integrated velocity ranges are $V_{LSR}$ = 1.0 - 17 km s$^{-1}$, 1.0 - 17 km s$^{-1}$, 3.0 - 17 km s$^{-1}$, for the $^{12}$CO, $^{13}$CO, and C$^{18}$O integrated intensity maps, respectively. The excitation temperature ($T_{\rm ex}$) is derived using the following equation \citep{Pineda08},

\begin{equation}   T_{\rm ex}=5.53\biggl\{\ln\biggl[1+\frac{5.53}{T_{\rm peak}+0.84}\biggr]\biggr\}^{-1},
   \label{eq:Tex}
\end{equation}
 where  $T_{\rm{peak}}$ is a peak brightness temperature of $^{12}$CO.
$^{12}$CO $T_{\rm ex}$ is adopted as one of the features because it should reflect the gas temperature in the LTE condition, which should be the dust temperature under the condition of the thermal coupling between gas and dust.

The maps of the intensity ratios are included as features to account for the influence of far ultraviolet (FUV) radiation.
The Orion A region contains a photon-dominated region (PDR)
where isotope selective photodissociation actively
occurs \citep{Shimajiri14_13CO/C18O, Ishii19_13CO/C18O, Lin16_13CO/C18O}. This causes variations in the intensity ratio among $^{12}$CO, $^{13}$CO, and C$^{18}$O.
To avoid the influence of noise, the ratio maps are made using pixels where the peak intensity for each line exceeds 5$\sigma$.
As a result, there are missing pixels of those ratios (see Figures \ref{fig0} (g) and (h)). The mapping regions of the 45-m observations are also different for the different CO isotopologue lines.
For this reason, a small fraction of pixels lack values for
a certain set of the molecular-line features.
PyCaret fills these missing data values with the relevant mean values
within the specified training area. This mean imputation
for missing data ensures that
every pixel retains a fixed-length feature vector required by the regression models.
Except for the ratio maps, no restriction of the feature data above a certain detection level is adopted, and noisy data are also included as they are.
Inclusion of noisy pixels should also provide the machine learning with important information, as it can learn a non-linear relationship from the whole ensemble data.

As the $^{12}$CO, $^{13}$CO, and the C$^{18}$O mom 0 maps
are also adopted as features, the ratio maps are redundant in
traditional concept of astrophysical data analyses.
In machine learning, however, this sort of information
is not necessary regarded as redundancy.
The same $^{12}$CO mom 0 / $^{13}$CO mom 0 ratio gives
a range of $^{12}$CO mom 0 and $^{13}$CO mom 0 values.
The plots between the $^{12}$CO mom 0 (or $^{13}$CO mom 0) v.s.
$^{12}$CO mom 0 / $^{13}$CO mom 0 ratio do not show any correlation,
and the data points rather scatter.
The $^{13}$CO and C$^{18}$O data sets also show a similar behavior.
Machine learning adopts those features as independent ones
to construct a model.

The dust temperature map produced from the Herschel data is
used as the target values. In Figure \ref{figA} correlations
between the dust temperature and the nine adopted features 
are shown along with the relevant correlation coefficient.
The plots are rather scattered, and none of the single feature
is sufficient to predict the dust temperature.
Features such as $^{12}$CO mom 0 and $^{12}$CO / $^{13}$CO ratio
show higher correlation coefficients with the dust temperature
than the other features.
Machine learning will extract possible relations
between the feature data and the target values.

For each pixel in the training area, the nine feature values are concatenated into a feature vector, which is then paired with the corresponding target value. In this way, each pixel is treated as one independent training sample.
An example of the training area is indicated by a black box (see Fig. \ref{fig0} (j)). 
This training area is selected because it contains a wide range of dust temperatures and the Orion Nebula and Orion Bar, i.e., PDR.
We call this region Regressor-Region-A, which is regarded as the fiducial region.
Table \ref{tab:example_training_table} shows data values in the five arbitrary pixels in the training area as a benchmark. This training area consists of 7.40\% of the total 337495 pixels in the entire observed area.
We note that the selection of this fiducial training area is somewhat subjective without any quantitative measure for the selection, such as stratified sampling across temperature bins.
In a realistic situation, however, measurements of dust temperatures are often available only in part of the target field of interest. In such cases, all the available dust temperature data in that sub-region should be used to predict the dust temperatures in the remaining area.
Our experiment is intended to simulate this practical situation.

\subsection{Selection of the Regression Model} \label{subsec: Selection_of_the_Regression_Model}

For our experiment, PyCaret version 2.3.10\footnote{\url{https://github.com/pycaret/pycaret}} with scikit-learn 0.23.2 \citep{Pedregosa2011} and scikit-optimize 0.8.1\footnote{\url{https://github.com/scikit-optimize/scikit-optimize}} in Python is adopted, which is a low-code machine learning framework.
PyCaret has 18 different types of machine-learning models, ``regression models'', which can be categorized into four groups.
Linear models (Linear Regression, Lasso Regression, Ridge Regression, Elastic Net, Least Angle Regression, Lasso Least Angle Regression, Bayesian Ridge, Huber Regressor, Orthogonal Matching Pursuit, Passive Aggressive Regressor) assume a linear relationship between features and targets.
Tree-based methods (Decision Tree, Extra Trees, Random Forest) partition the features and connect them by ``trees" to capture the non-linear relation to the targets.
Boosting algorithms (AdaBoost, Gradient Boosting, Light Gradient Boosting) sequentially combine weak learners to improve predictive accuracy. Other approaches include k-Nearest Neighbors and Dummy Regressor.

We first need to decide which regression model should be adopted for our specific case.
The features and target values in Regressor-Region A were given to the {\it compare\_models} function in PyCaret, which performs cross-validation of all the 18 regression models using their default hyperparameters.
In the cross-validation stage, Pycaret divides the  features and the target value dataset into {\it k} subsets of equal size without any overlap ({\it k} = 10).
{\it k}-1 folds are used as training data, while the remaining fold is used as validation data. Since the validation data is not used in the training, it can be used to evaluate the performance of the trained model.
This procedure is performed until all the {\it k} subsets are used as validation data, and the average statistical value of {\it k} rounds is compared among the 18 models to select the best model. Table \ref{tab:compare_models_table} compares Mean Absolute Error (MAE), Mean Square Error (MSE), Root Mean Square Error (RMSE), R-squared (R2), Root Mean Squared Logarithmic Error (RMSLE), Mean Absolute Percentage Error (MAPE), and TT (training time) for each model. Models with lower values for MAE, RMSE, RMSLE, and MAPE indicate better performance. On the other hand, a higher R2 value indicates a better model, which is defined as
\begin{equation}
    R2=1-\frac{\sum_{i=1}^n(\hat{y_i}-y_i)^2}{\sum_{i=1}^n(\bar{y}-y_i)^2},
	\label{eq:R2}
\end{equation}
where {\it n} is the total number of pixels in Regressor-Region A,
including both training and non-training areas, ${\hat{y_i}}$ is the predicted value for the {\it~i}~th pixel, ${y_i}$ is the observed value for the {\it~i}~th pixel, and ${\bar{y}}$ is the mean of the observed values.
The Extra Trees Regressor (ET) performs best across all the statistical values, and it is selected for use in this study (see Table \ref{tab:compare_models_table}).
We found that even in the case of other training areas, ET is always chosen as the best regressor model across all the statistical means.

ET is a regression model structurally similar to the Random Forest Regressor (RF), as both are ensemble methods built upon decision trees. In a decision tree, each node selects a feature, each branch represents a splitting condition, and each leaf corresponds to the final regression output. However, a single tree often fits the training data too closely, leading to high variance and overfitting.
The RF algorithm mitigates this problem by training multiple trees on different subsets of the data, where each subset is generated randomly, allowing duplication of the same data. This data sampling process is called ``bootstrapping".
This resampling technique allows each tree to learn slightly different patterns from the data.
Averaging the predictions from these diverse trees reduces variance and improves generalization performance.
On the other hand, ET does not adopt bootstrapping but a different randomness.
Unlike RF, ET uses the entire dataset and randomly selects the split thresholds from the range of possible feature values instead of searching for the optimal threshold.
This additional source of randomness increases tree diversity and makes the ensemble more robust against overfitting.
For our dataset, which contains correlated physical quantities and a limited number of independent samples, ET performed better than RF.

\begin{table*}
    \centering
    \vspace{-1em}
    \caption{Comparison among models}
    \label{tab:compare_models_table}
    \scriptsize
    \begin{tblr}{
        colspec = {l l r r r r r r r},
        colsep = 2pt,
        rowsep = 0.5pt,
    }
    \hline
    & Model & MAE & MSE & RMSE & R2 & RMSLE & MAPE & TT \\
    & & \emph{Mean Absolute} & \emph{Mean Square} & \emph{Root Mean} & \emph{R-square} & Root Mean & \emph{Mean Absolute} & \emph{Training} \\
    & & \emph{Error} & \emph{Error} & \emph{Square Error} & & \emph{Square Log Error} & \emph{Percentage Error} & \emph{Time} \\
    \hline
    & & [K] & [K$^2$] & [K] & & & [\%] & [sec] \\
    \hline
    et & Extra Trees Regressor & 0.3739 & 0.4753 & 0.6890 & 0.9901 & 0.0254 & 0.0154 & 0.2870 \\
    rf & Random Forest Regressor & 0.4444 & 0.6776 & 0.8222 & 0.9859 & 0.0297 & 0.0182 & 0.6970 \\
    lightgbm & Light Gradient Boosting & 0.6791 & 1.1223 & 1.0592 & 0.9766 & 0.0397 & 0.0292 & 0.0290 \\
    dt & Decision Tree Regressor & 0.5368 & 1.4189 & 1.1893 & 0.9704 & 0.0429 & 0.0221 & 0.0440 \\
    knn & K Neighbors Regressor & 0.5885 & 1.4202 & 1.1909 & 0.9704 & 0.0407 & 0.0232 & 0.0220 \\
    gbr & Gradient Boosting Regressor & 0.9415 & 2.0470 & 1.4300 & 0.9574 & 0.0532 & 0.0401 & 0.8850 \\
    ada & AdaBoost Regressor & 1.4669 & 3.5725 & 1.8894 & 0.9256 & 0.0775 & 0.0686 & 0.2390 \\
    ridge & Ridge Regression & 1.7877 & 6.5512 & 2.5589 & 0.8637 & 0.0950 & 0.0772 & 0.0070 \\
    br & Bayesian Ridge & 1.7876 & 6.5512 & 2.5589 & 0.8637 & 0.0950 & 0.0772 & 0.0140 \\
    lr & Linear Regression & 1.7877 & 6.5512 & 2.5589 & 0.8637 & 0.0950 & 0.0772 & 0.1390 \\
    lar & Least Angle Regression & 1.8045 & 6.7162 & 2.5908 & 0.8603 & 0.0964 & 0.0780 & 0.0080 \\
    huber & Huber Regressor & 1.7669 & 7.0744 & 2.6590 & 0.8529 & 0.0968 & 0.0738 & 0.0880 \\
    en & Elastic Net & 1.9283 & 7.1058 & 2.6650 & 0.8521 & 0.1019 & 0.0859 & 0.0180 \\
    lasso & Lasso Regression & 1.9332 & 7.1575 & 2.6747 & 0.8511 & 0.1019 & 0.0859 & 0.0140 \\
    omp & Orthogonal Matching Pursuit & 2.6116 & 17.5130 & 4.1832 & 0.6361 & 0.1405 & 0.1039 & 0.0060 \\
    par & Passive Aggressive Regressor & 3.2999 & 26.3100 & 4.9207 & 0.4544 & 0.1888 & 0.1370 & 0.0180 \\
    llar & Lasso Least Angle Regression & 5.5282 & 48.1331 & 6.9365 & -0.0006 & 0.2763 & 0.2552 & 0.0060 \\
    dummy & Dummy Regressor & 5.5282 & 48.1331 & 6.9365 & -0.0006 & 0.2763 & 0.2552 & 0.0070 \\
    \hline
    \end{tblr}
    \par
    \vspace{0.5em}
    \raggedright \footnotesize Comparison of regression models using the \textit{compare\_models} function, with Regressor-Region-A as the training area.
\end{table*}

\begin{table*}
    \centering
    \vspace{1em}
    \begin{talltblr}[
        caption = {Tuned hyperparameters of ET for each training area and prediction results},
        label = {tab:tuned_dataset_table},
        note{\textdagger} = {: RA and DEC coordinates of the bottom left corner of the training area.},
        note{\textdaggerdbl} = {: RA and DEC coordinates of the top right corner of the training area.},
        note{1} = {: Maximum depth of the tree.},
        note{2} = {: Limit of the maximum number of features used for each tree.},
        note{3} = {: Minimum number of samples required to be at a leaf node.},
        note{4} = {: Minimum number of samples required to split an internal node.},
        note{5} = {: Minimum weighted fraction of the sum total of weights required to be at a leaf node.},
        note{6} = {: Number of trees in the forest.},
        note{7} = {: Number of selected features in ET.}
      ]{
        colspec = {l c c c c c c},
        colsep = 2pt,
        rowsep = 1.5pt,
        width = \linewidth
      }
        \hline
        Area used for training & Regressor- & Regressor- & Regressor- & Regressor- & Regressor- & Regressor- \\
        & Region-A & Region-B & Region-C & Region-D & Region-E & Region-F \\
        \hline
        blc\_RA\TblrNote{\textdagger} & 5$^{\rm h}$35$^{\rm m}$43$^{\rm s}$.7 & 5$^{\rm h}$36$^{\rm m}$08$^{\rm s}$.8 & 5$^{\rm h}$37$^{\rm m}$14$^{\rm s}$.4 & 5$^{\rm h}$36$^{\rm m}$33$^{\rm s}$.9 & 5$^{\rm h}$34$^{\rm m}$43$^{\rm s}$.4 & 5$^{\rm h}$35$^{\rm m}$43$^{\rm s}$.7 \\
        blc\_DEC\TblrNote{\textdagger} & -6$^{\rm d}$23$^{\rm m}$28$^{\rm s}$.8 & -5$^{\rm d}$34$^{\rm m}$43$^{\rm s}$.7 & -6$^{\rm d}$44$^{\rm m}$43$^{\rm s}$.0 & -5$^{\rm d}$32$^{\rm m}$13$^{\rm s}$.5 & -5$^{\rm d}$15$^{\rm m}$58$^{\rm s}$.8 & -6$^{\rm d}$23$^{\rm m}$28$^{\rm s}$.8 \\
        trc\_RA\TblrNote{\textdaggerdbl} & 5$^{\rm h}$35$^{\rm m}$19$^{\rm s}$.0 & 5$^{\rm h}$34$^{\rm m}$53$^{\rm s}$.9 & 5$^{\rm h}$35$^{\rm m}$29$^{\rm s}$.1 & 5$^{\rm h}$33$^{\rm m}$38$^{\rm s}$.6 & 5$^{\rm h}$33$^{\rm m}$38$^{\rm s}$.7 & 5$^{\rm h}$35$^{\rm m}$19$^{\rm s}$.0 \\
        trc\_DEC\TblrNote{\textdaggerdbl} & -5$^{\rm d}$21$^{\rm m}$06$^{\rm s}$.4 & -5$^{\rm d}$12$^{\rm m}$21$^{\rm s}$.4 & -6$^{\rm d}$28$^{\rm m}$36$^{\rm s}$.4 & -5$^{\rm d}$22$^{\rm m}$21$^{\rm s}$.0 & -4$^{\rm d}$49$^{\rm m}$51$^{\rm s}$.0 & -5$^{\rm d}$14$^{\rm m}$51$^{\rm s}$.4 \\
        \hline
        Number of samples (pixels) & 25000 & 27000 & 27300 & 28000 & 27300 & 27500 \\
        in training area & 7.40\% & 8.00\% & 8.08\% & 8.29\% & 8.08\% & 8.14\% \\
        \hline
        max\_depth\TblrNote{1} & 11 & 11 & 11 & 11 & 11 & 11 \\
        max\_features\TblrNote{2} & 1 & 1 & 1 & 1 & 1 & 1 \\
        min\_samples\_leaf\TblrNote{3} & 1 & 1 & 1 & 1 & 1 & 1 \\
        min\_samples\_split\TblrNote{4} & 2 & 2 & 2 & 2 & 2 & 2 \\
        min\_weight\_fraction\_leaf\TblrNote{5} & 0 & 0 & 0 & 0 & 0 & 0 \\
        n\_estimators\TblrNote{6} & 300 & 158 & 300 & 203 & 248 & 296 \\
        Number of features selected\TblrNote{7} & 8 & 8 & 8 & 8 & 8 & 8 \\
        \hline
        R2 & 0.9767 & 0.7758 & 0.9042 & 0.9095 & 0.8531 & 0.9564 \\
        Average temperature & 20.90$\pm$4.51 & 27.71$\pm$2.55 & 16.48$\pm$1.11 & 23.92$\pm$3.58 & 19.68$\pm$1.94 & 21.35$\pm$5.01 \\
        \hline
    \end{talltblr}
\end{table*}

\subsection{Optimization of Set-up} \label{sub: Optimization_of_Set-up}
Tuning of the hyperparameters of the ET regressor for different training areas should then be performed. The adopted hyperparameters for each training area are listed in the fourth row of Table \ref{tab:tuned_dataset_table}.
The ${\it tune\_model}$ function in the PyCaret module can automatically search for the optimal set of hyperparameters. ${\it tune\_model}$ performs a Bayesian search to minimize RMSE in each training area, and over the 500 iterations, the most optimal set of
the hyperparameters is derived. Besides, to evaluate the importance of the features in constructing a decision tree and to determine which features should be adopted, Recursive Feature Elimination with Cross-Validation (RFECV) is performed. Recursive Feature Elimination (RFE) eliminates the least important features.
This process helps eliminate irrelevant or redundant features, thereby reducing model complexity and improving generalization performance.
In cross-validation, the training data and validation data are evaluated separately. The score (R2) obtained from the training data is called the training score, while the score from the validation data is called the cross-validation score. 
 
\begin{figure*}[htbp]
\centering
\includegraphics[width=17cm]{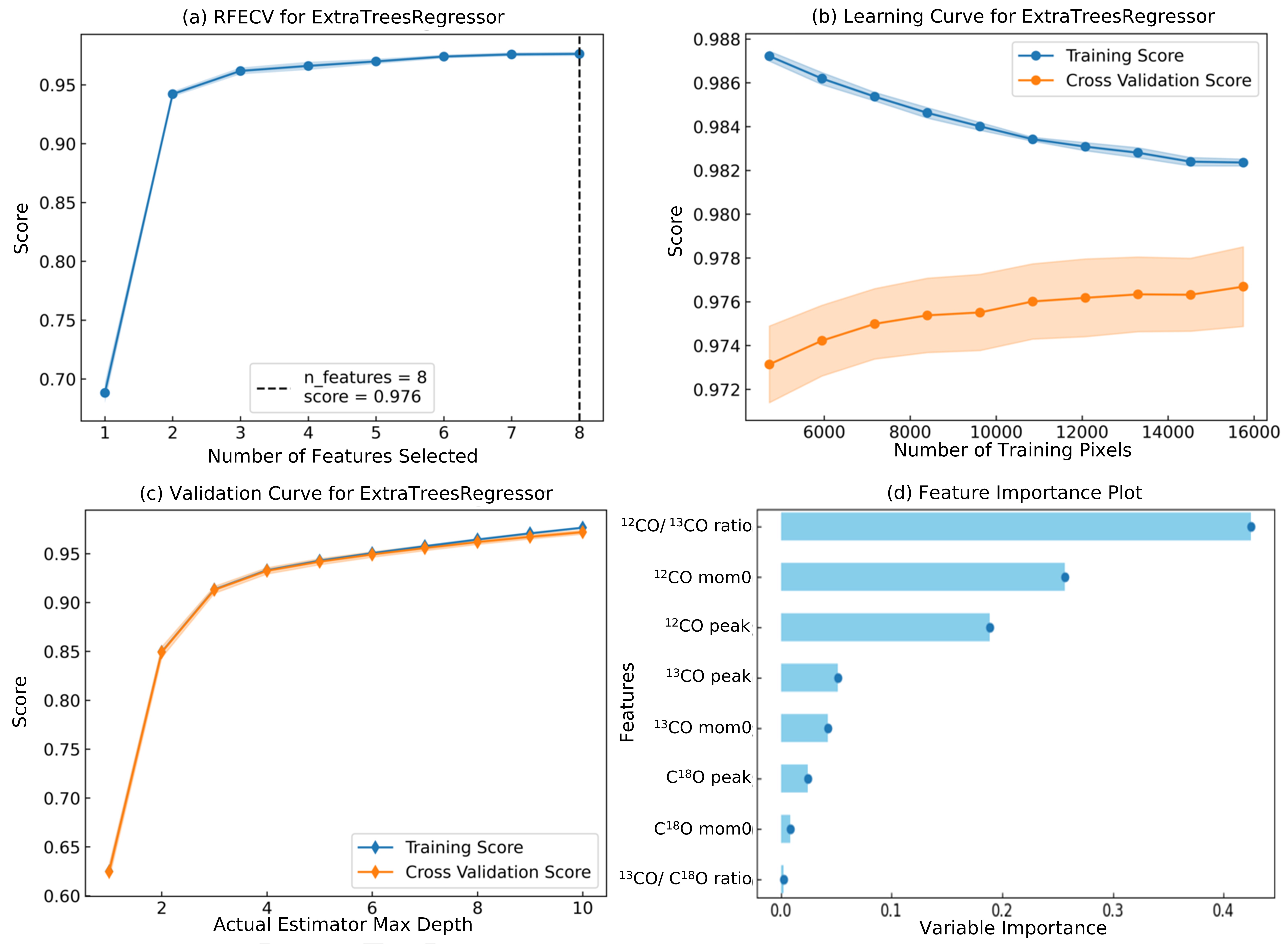}
\caption{Plots of (a) recursive feature elimination with cross-validation (RFECV),
(b) training and cross-validation scores as a function of the
number of the training pixels, i.e., learning curves,
(c) training and validation curves as a function of the maximum depth of the tree,
and (d) feature importance, for Regressor-Region-A.
In panels (b) and (c), the blue and orange curves show the training and cross-validation scores, respectively, with the range of the standard deviation. Panel (a) shows only the cross-validation scores, along with their standard deviations.
The scores refer to R2.}
\label{fig1}
\end{figure*}

Figure \ref{fig1} presents various output plots from the ${\it tune\_model}$ function for Regressor-Region-A. Figure \ref{fig1}(a) shows the cross-validation score as a function of the number of adopted features and demonstrates how the number of features is determined through RFECV. The optimal number of features is 8.
In this model, $^{12}$CO $T_{\rm ex}$ is considered redundant and eliminated.
$^{12}$CO $T_{\rm ex}$ is calculated from the $^{12}$CO peak brightness temperature with equation 1, and thus the plots between $T_{dust}$ and the $^{12}$CO peak brightness temperature and between $T_{dust}$ and $^{12}$CO $T_{\rm ex}$ are identical (Fig. \ref{figA}).
${\it tune\_model}$ detected such a correspondence and eliminated one of them.
Figure \ref{fig1}(b) shows the training and validation scores as a function of the number of training pixels, including the standard deviation for different cross-validation within regressor-region-A. 
If the training score is significantly higher than the cross-validation score, it may indicate that the regressor model is overfitting the training data.
Closer scores from both the training and validation data indicate a better model. 
These curves demonstrate how the model gains better learning with an increasing number of training pixels, $i.e.,$ learning curve.
Figure \ref{fig1}(c) shows the dependence of the training and validation scores as a function of the maximum depth of the decision trees. The model performance improves rapidly up to a tree depth of 4, after which the improvement becomes more gradual. The final optimal depth is determined based on RMSE; since slight improvements continue beyond depth 4, a moderately deeper tree is selected as the optimal configuration. The dependence is almost identical in both the training and cross-validation scores, suggesting that the overfitting is successfully avoided.
After the validation process of the hyperparameters, the importance of each feature can be derived. Figure \ref{fig1}(d) shows the percentage of
the number of the decision-tree splits which adopt that specific feature
in regressor-region-A. This percentage can be regarded as a relative
importance of that particular feature in the model.
As described below, in most cases, the $^{12}$CO / $^{13}$CO ratio turns out to be the most important feature in determining the dust temperature.

\begin{figure*}[htbp]
\centering
\includegraphics[width=13cm]{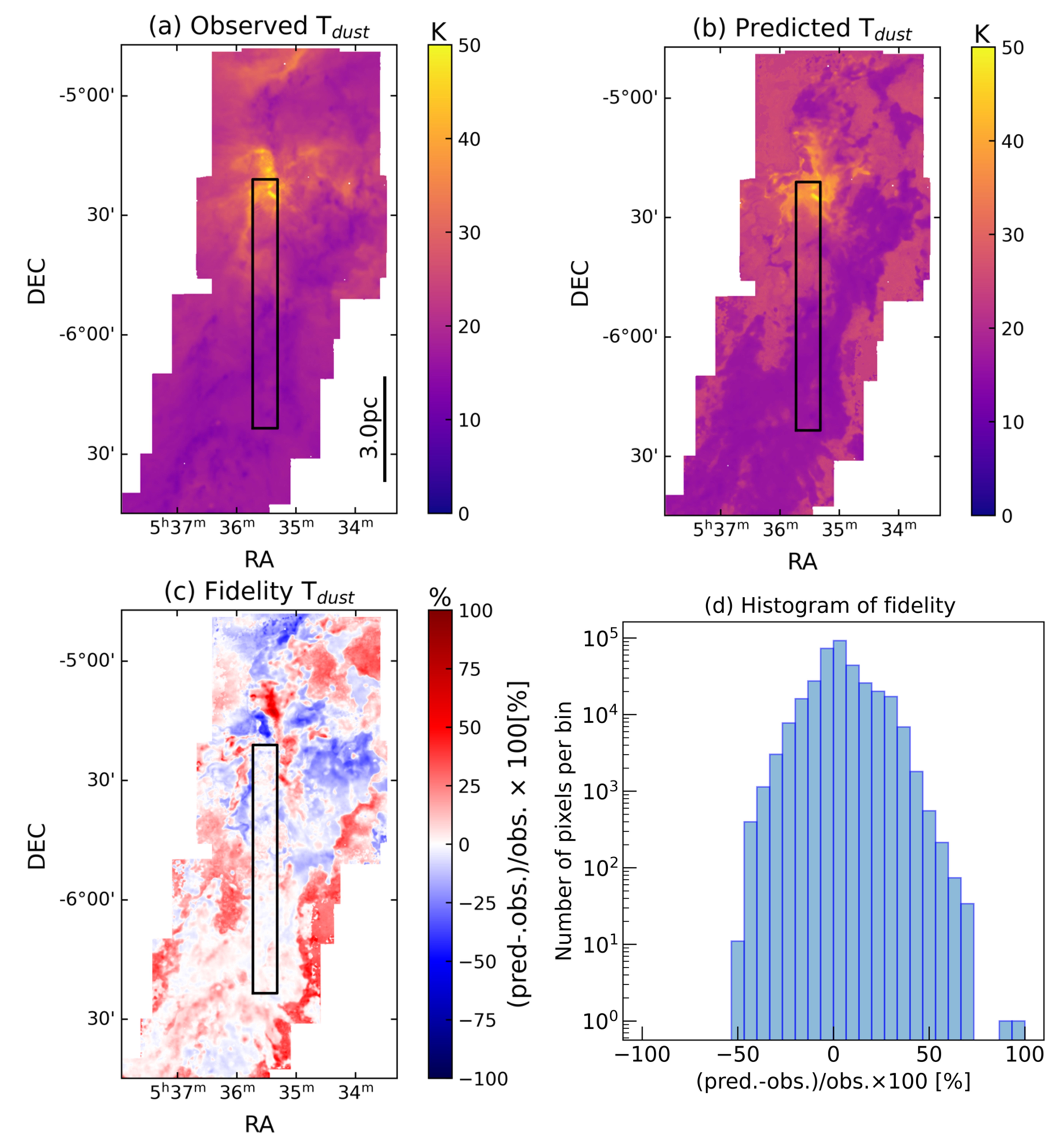}
\caption{Maps of (a) the dust temperature in the Orion A region produced from the HGBS data, (b) the dust temperature predicted from Regressor-Region-A, (c) the fidelity between the observation and prediction, and (d) the histogram of the fidelity. The black boxes in panels (a), (b), and (c) show the area used for training the regressor.}
\label{fig2}
\end{figure*}

\section{RESULTS} \label{sec: results}

Figures \ref{fig2}(a) and (b) compare the observed ($T_{\rm dust, obs}$) and predicted dust temperature ($T_{\rm dust, pre}$) maps with Regressor-Region A as the training area
(black rectangles in Figures \ref{fig2} (a), (b) and (c)).
Overall, the predicted dust temperature map reproduces the true, observed dust temperature map, such as the hot regions around Orion KL and the extended low-temperature regions to the south. 
The predicted map, however, exhibits sharper structures and image contrast, even though the overall shapes of the structure resemble those of the real map.
This sharpness in the predicted map can be understood from the nature of the ET regressor.
Unlike deep learning methods or other algorithms that approximate smooth continuous functions, ET partitions the feature space into discrete intervals defined by multiple thresholds. Even a small change in input features can result in a jump to a different leaf, producing abrupt transitions in the output. Depending on the dataset, ET may emphasize spatially discrete structures.

The map of the difference between the predicted and observed dust temperatures, normalized by the observed dust temperatures, which we call ``fidelity", presents a more quantitative difference of the prediction from the target values (Figure \ref{fig2}(c)).
There are trends that the predicted dust temperature is underestimated in the northern regions and overestimated in the southern regions. 
Figure \ref{fig2}(d), the histogram of fidelity, shows that 61.83\% of predicted pixels have positive fidelity, while 38.17\% have negative fidelity. 90\% of all pixels fall within the range of -17.72\% to 30.08\%. The pixels with overestimated values are more frequent compared to those with underestimated values. 

\begin{figure*}[htbp]
\centering
\includegraphics[width=16cm]{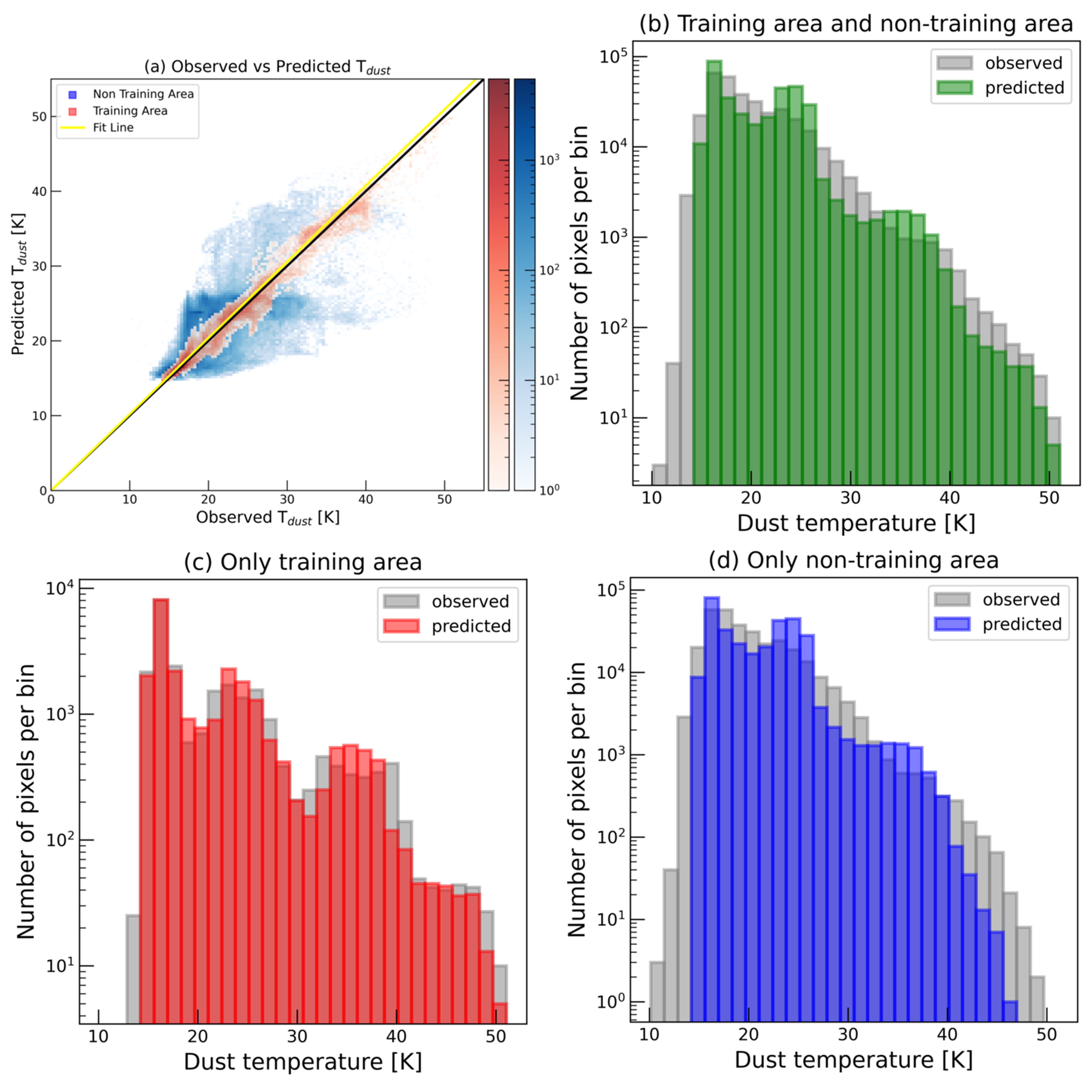}
\caption{(a) Pixel-to-pixel correlation between the predicted and observed dust temperatures. Red points denote data points from the training area, while blue points outside the training area. Opacity of the colors indicates the density of the data points.
The black line shows the identical predicted and observed dust temperatures, and the yellow line shows the relationship $T_{\rm dust, pre}$ = (1.016 $\pm$ 0.001) $\times$ $T_{\rm dust, obs}$ obtained through the least-square fitting.
(b) Pixel-to-pixel histograms of the predicted (color) and observed (gray) dust temperature in the whole Orion A region, (c) those within the area of Regressor-Region-A (i.e., inside the black box in Fig. \ref{fig2}(a)), and (d) those outside Regressor-Region-A.}
\label{fig3}
\end{figure*}

Figure \ref{fig3} shows various comparisons between the observed and predicted dust temperatures. The correlation plot between the observed and predicted dust temperatures, shown in Figure \ref{fig3}(a), demonstrates that the predicted dust temperatures decently reproduce the true observed dust temperatures. It is natural that the predicted dust temperatures within the training area, shown in red points, better reproduce the observed dust temperatures than those outside the training area (blue points). The least-square fitting to all the data points yields the correlation of $T_{\rm dust, pre}$ = (1.016 $\pm$ 0.001) $\times$ $T_{\rm dust, obs}$, an almost perfect linear relation between the predicted and observed dust temperatures. Figures \ref{fig3}(b), (c), and (d) show histograms of the observed and predicted dust temperatures in the entire area, within the training area, and outside the training area, respectively. Overall, the statistical distributions of the real and predicted dust temperatures agree well. As already described above, the statistics of the predicted dust temperatures within the training area better match that of the observed ones (Fig. \ref{fig3}(c)) than outside the training area (Fig. \ref{fig3}(d)). Specifically, since the training area does not include dust temperatures below 14 K, the regressor model cannot predict such low temperatures, as clearly seen in Figures \ref{fig3}(b) and (d).  The prediction accuracy is also low at above 45 K. In the training area, the fraction of the observed pixels with above 45 K reaches 0.55\%, while that outside the training area is only 0.016\%.
This imbalance in the representation of high-temperature pixels leads to the underestimation of such pixels in regions outside the training area.
These results imply that a more diverse and widely covered sample of the dust temperature in the training is essential to improve the accuracy of the machine-learning prediction.

In summary, the average observed dust temperature is 20.25 K with a standard deviation of 4.60 K, while the average predicted dust temperature is 20.87 K with a standard deviation of 4.48 K. This suggests that Regressor-Region-A predicts the dust temperatures from the molecular-line data well. However, the prediction accuracy becomes lower for the dust temperature ranges that are not properly sampled in the training area.

\section{DISCUSSION} \label{sec: discussion}

\begin{figure*}[htbp]
\centering
\includegraphics[width=18cm]{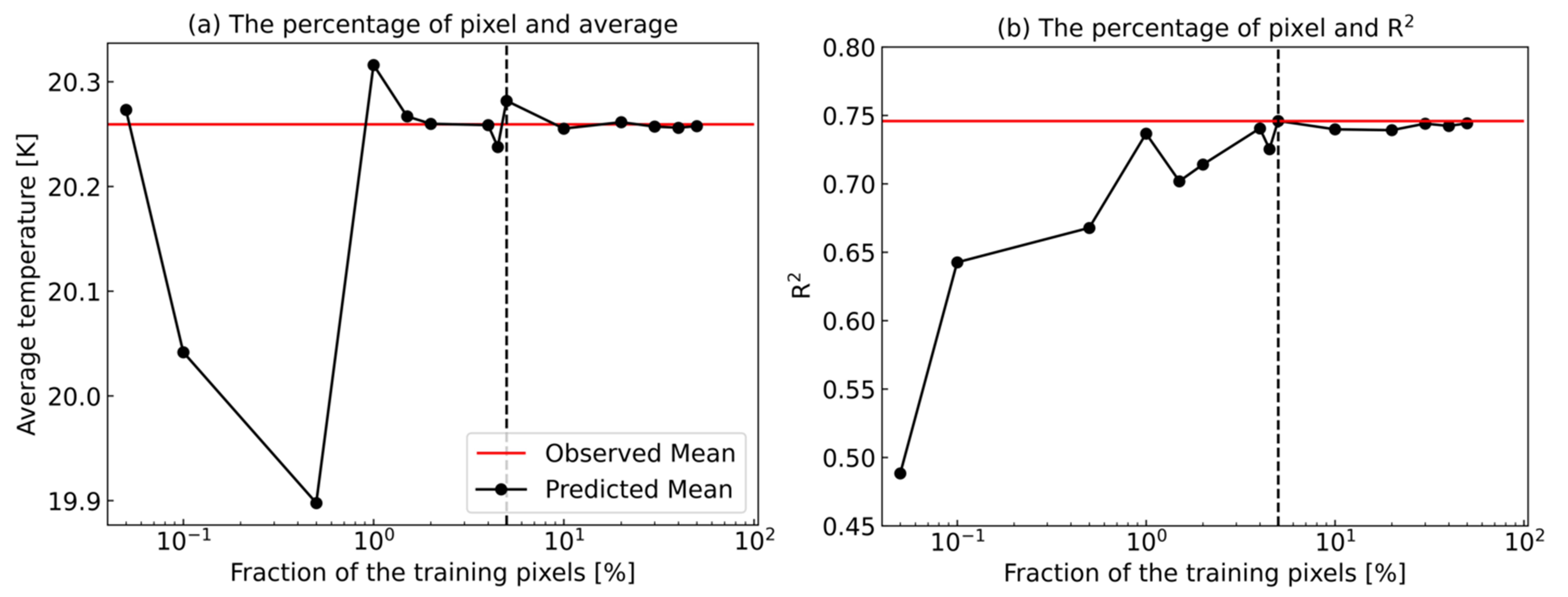}
\caption{Relationships (a) between the fraction of the training pixels and the average predicted dust temperature and (b) between the fraction of the training pixels and R2. The predicted map is generated using a regressor trained on a randomly selected subset of pixels, covering a specific percentage of the total pixels. The red line in panel (a) indicates the average observed dust temperature. The dotted lines in panels (a) and (b) indicate the fraction of the training pixels of 5\% of the entire pixels, which is regarded as a sufficient level of pixel number for accurate prediction. At this point, the R2 value is 0.7409 (red line in panel b).}
\label{fig4}
\end{figure*}

\begin{figure*}[htbp]
\includegraphics[width=18cm]{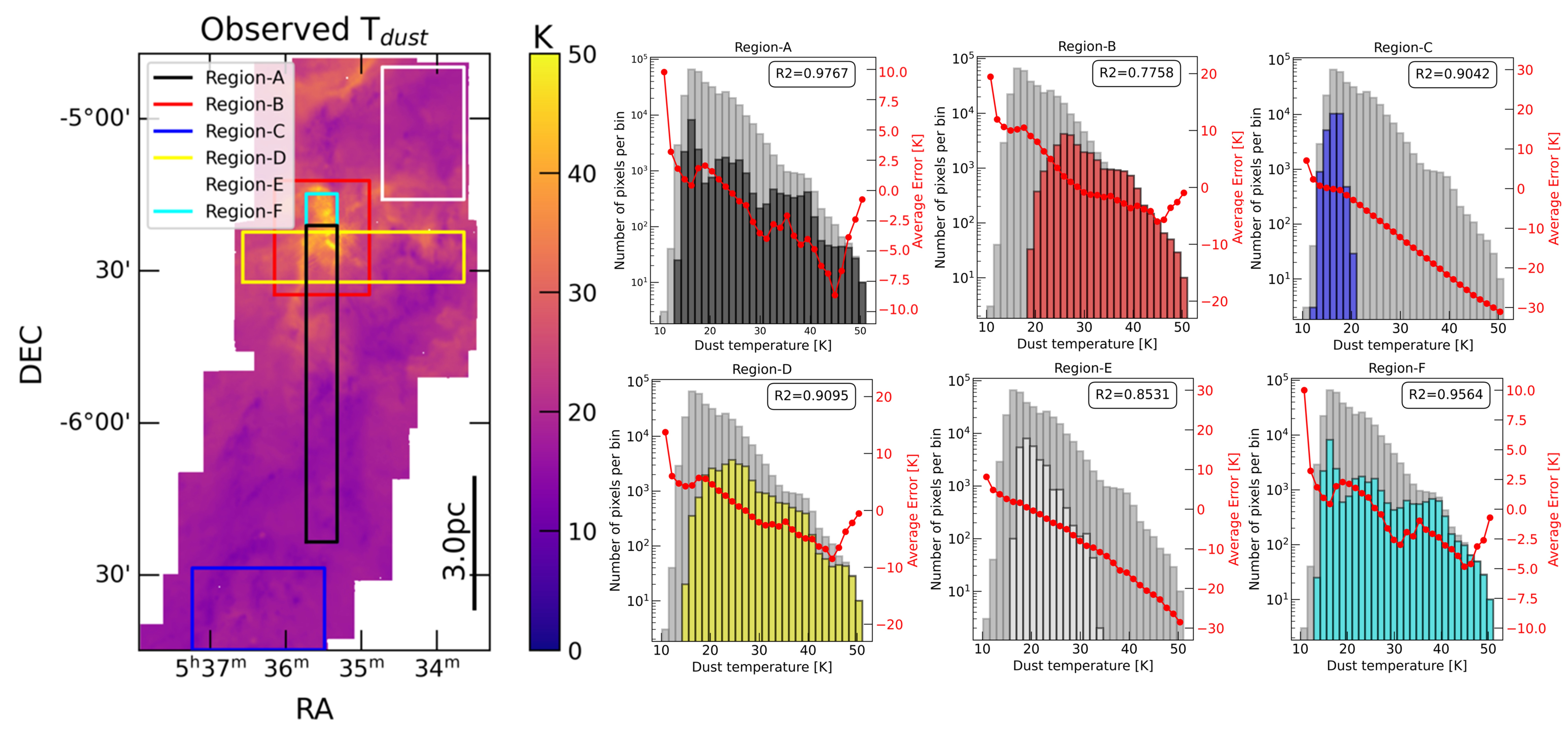}
\caption{(Left) Locations of the six training areas on the observed dust temperature map.
Black, red, blue, yellow, white, and cyan boxes indicate the area of Regressor-Region A, B, C, D, E, and F, respectively.
(Right) Corresponding histograms of the observed dust temperature in the entire (gray) and the relevant training areas (color).
The resultant R2 values using the training area are also written.
In the histograms, the mean absolute errors of the predicted values in each temperature bin are plotted (red point).}
\label{fig5}
\end{figure*}

\begin{table*}[htbp]
  \tiny
  \centering
  \caption{Characteristics and the Most Important Features in Each Training Area}
  \label{tab:table4}
  \begin{tblr}{
  colspec = {l c c c c c c c c c c c c c},
  hlines,
  vlines,
  row{1,2} = {font=\bfseries},
}

\SetCell[r=2]{c} Region
& \SetCell[c=3]{c} Region Type
&&& \SetCell[r=2]{c} $T_{\rm d,min}$
& \SetCell[r=2]{c} $T_{\rm d,max}$
& \SetCell[r=2]{c} $T_{\rm d,mean}$
& \SetCell[r=2]{c} $T_{\rm d,std}$
& \SetCell[r=2]{c} $A_{\rm V,min}$
& \SetCell[r=2]{c} $A_{\rm V,max}$
& \SetCell[r=2]{c} $A_{\rm V,mean}$
& \SetCell[r=2]{c} $A_{\rm V,std}$
& \SetCell[r=2]{c} Most Important Feature \\

& Low-$T_{\rm d}$
& High-$T_{\rm d}$
& PDR
& & & & & & & & & \\

Region-A & N & Y & Y & 14.03 & 51.05 & 21.49 & 6.92 & 1.47 & 84.99 & 6.56 & 4.99 & $^{12}$CO/$^{13}$CO \\
Region-B & Y & N & Y & 19.26 & 51.05 & 29.25 & 5.23 & 1.01 & 51.04 & 13.24 & 29.53 & $^{12}$CO/$^{13}$CO \\
Region-C & Y & N & N & 12.85 & 20.14 & 16.48 & 1.11 & 1.71 & 37.92 & 5.39 & 3.50 & $^{12}$CO peak \\
Region-D & Y & Y & N & 15.15 & 51.05 & 25.70 & 5.54 & 1.42 & 35.04 & 8.37 & 22.82 & $^{12}$CO/$^{13}$CO \\
Region-E & Y & N & N & 16.23 & 31.43 & 20.38 & 2.53 & 0.90 & 11.82 & 2.71 & 1.45 & $^{12}$CO mom0 \\
Region-F & Y & Y & Y & 14.03 & 51.05 & 22.76 & 7.85 & 1.47 & 109.75 & 7.08 & 6.46 & $^{12}$CO/$^{13}$CO \\
\end{tblr}
\end{table*}
 
\subsection{Influence of the pixel number in the training area on the prediction accuracy} \label{sec: fraction}

To investigate the effect of the number of pixels for the training on the prediction accuracy, the number of pixels in the training area is varied from 0.05 \% to 50 \% of the entire observed region (337495 pixels). The training pixels are selected by
random sampling over the entire map and thus
spatially scattered and non-contiguous.
In the 0.05 \% case, PyCaret chooses Random Forest as the most optimal regressor model instead of ET, while in all the other cases, the ET regressor is chosen.
Figure \ref{fig4}(a) shows the relation between the percentage of the training pixels and the average predicted dust temperature. The variation of the average predicted dust temperature from 19.89 K to 20.31 K is seen for the fraction of the training pixels up to 5 \%, while the variation becomes much smaller (from 20.25 K to 20.28 K) for fractions of 5 \% or above.
The average predicted dust temperature at the 0.5 \% pixel number shows the biggest difference from the observed dust temperature. Even in this case, the difference between the average observed and predicted dust temperatures is 1.78 \%, much smaller than the standard deviation of the observed and predicted dust temperatures in the entire observed region (20.25$\pm$4.60 K and 19.89$\pm$3.64 K).
Thus, even if a fraction of the training pixels is as small as ~0.5 \% or so, we can still achieve a reasonable prediction accuracy.

Figure \ref{fig4}(b) shows R2 as a function of the fraction of the training pixels. Above 5 \% of the training pixels, the R2 value becomes almost saturated.
To look into this result in more detail, Figure \ref{figB} lists the predicted dust temperature and fidelity maps, histograms of the dust temperatures of the entire and training pixels, and the feature importance plots in the case of the 0.05, 0.1, 5, and 20 \% training pixels.
The R2 value is $\sim$0.49 in the 0.05 \% case.
The lack of high ($>$40 K) and low temperatures ($<$14 K) in this small number of training pixels limits the prediction accuracy at those extreme temperatures.
This is seen in the high absolute value of the average error at the low- and high temperatures (red points in Figure \ref{figB}), as well as in the highest amplitude of the fidelity image.
The R2 value ($\sim$0.64) and fidelity map slightly improve in the 0.1 \% case, but these training pixels still do not have a high-temperature regime above 40 K. With the fraction of the training pixels higher than 5 \%, the R2 value saturates, and there is no discernible change of the predicted and fidelity maps between the 5 \% and 20 \% pixel cases.
The histograms of the training pixels adequately sample the temperature range, and their shapes closely match that of the overall histogram. The absolute values of the average error become smaller, and there is less variation of the average error as a function of the temperature bins.
Furthermore, the feature ranking is now identical between the 5 \% and 20 \% cases, where the $^{12}$CO / $^{13}$CO ratio and the $^{12}$CO peak intensity are the primary and secondary important features, respectively.
These results show that once a well-sampled temperature coverage is provided to the machine learning, the regressor model becomes robust enough to provide consistent predictions.

\subsection{Dependence of prediction accuracy on the selection of the training area} \label{sec: box}

In the last subsection, the dependence of the prediction accuracy on the number of training pixels is discussed. In this subsection, we will also investigate the dependence on the selection of the training area.
Figure \ref{fig5} shows the selected six training areas and the histograms of the observed dust temperatures in the training (colored) and entire regions (gray).
The results of Regressor-Region-A are already described above.
Regressor-Region-B focuses on high-temperature regions without low-temperature (below 19 K) pixels. This training area includes PDRs such as the M43 shell and the Orion Bar \citep{Shimajiri14_13CO/C18O}. Regressor-Region-C is characterized by low temperatures (below 20 K) and relatively weak molecular lines, located in the southernmost part of the entire region. Regressor-Region-D encompasses a horizontally elongated region covering a wide temperature range from 15 to 51 K. Regressor-Region-E encloses a relatively low-temperature area to the northwest. Finally, Regressor-Region-F is an extension of Regressor-Region-A and comprises both Regressor-Region-A and the northern extension as delineated by the light blue box in Figure \ref{fig5} (left).
All of these training areas contain more than 5\% of the total number of pixels in the entire region, thus satisfying the criterion discussed in the last subsection.
The ET regressor was selected as the most optimal regressor for all six training areas, and the optimal hyperparameters for each region are summarized in Table \ref{tab:tuned_dataset_table}.
Table \ref{tab:table4} summarizes astrophysical quantities in these training areas.

The predicted dust temperature and fidelity maps, and the feature importance plot for each regression region are shown in Figures \ref{figC1} - \ref{figC6}.
Table \ref{tab:tuned_dataset_table} lists the resultant R2 values and the average predicted temperatures with these training areas.
After all, Regressor-Region-A, our fiducial training area, gives the highest R2 value.
By contrast, Regressor-Regions B and C, which are biased toward high and low dust temperatures (see Figure \ref{fig5} right), prefer to predict the high and low dust temperatures in the entire region, respectively (Figures \ref{figC2} and \ref{figC3}). 
Plots of the average error show a large positive error in the lower-temperature bin for Regressor-Region-B and a large negative error in the higher-temperature bin for Regressor-Region-C.
The most important feature to determine the dust temperature is also different, and Regressor-Region-C takes the $^{12}$CO peak as the most important feature in contrast to the $^{12}$CO / $^{13}$CO ratio adopted by Regressor-Region-A and B. Regressor-Region-E, which is also biased to lower dust temperatures, shows a similar trend to that of Regressor-Region-C to a lesser degree (Figure \ref{figC5}).
Regressor-Region-E contains slightly higher temperature pixels than Regressor-Region-C, which improves the prediction accuracy.

Whereas Regressor-Region-D has both high- and low-temperature pixels, the sample is slightly skewed toward a higher temperature as compared to that of Regressor-Region-A. 
The overall shape of the sample of Regressor-Region-D is also different from that of the entire observing region.
This slight skewness of the training sample makes the predicted temperatures higher than the observed values (Figure \ref{figC4}).
Regressor-Region-F is an extension of the fiducial training area of Regressor-Region-A. The predictions from Regressor-Region-F (Figure \ref{figC6}) are as accurate as those of the fiducial training area.
Note that there is no improvement in the prediction by the northern extension of the training area in Regressor-Region-F.
These results indicate that a more balanced sample of the dust temperatures in the training area, similar to that of the entire observing region, is essential to obtain more accurate predictions. 
Simple addition of some of the training pixels does not improve the prediction.

We also note that the most important feature in the case of Regressor-Region-A and F, which gives the most accurate predictions, is the $^{12}$CO/$^{13}$CO ratio.
These regressors are trained on areas that include prominent PDRs, where FUV photons dominate the energy balance or chemistry of the gas.
The FUV radiation dissociates rarer CO isotopologues more than the more abundant isotopologues because of the difference in self-shielding, i.e., selective isotope dissociation.
Therefore, the variation in the $^{12}$CO / $^{13}$CO ratio is created by the FUV radiation, which likely influences the dust temperature. 

\subsection{Effect of Gas-Dust Thermal Coupling on Dust Temperature Prediction from Molecular Line Data} \label{sec: Density}

\begin{figure*}[htbp]
    \includegraphics[width=18cm]{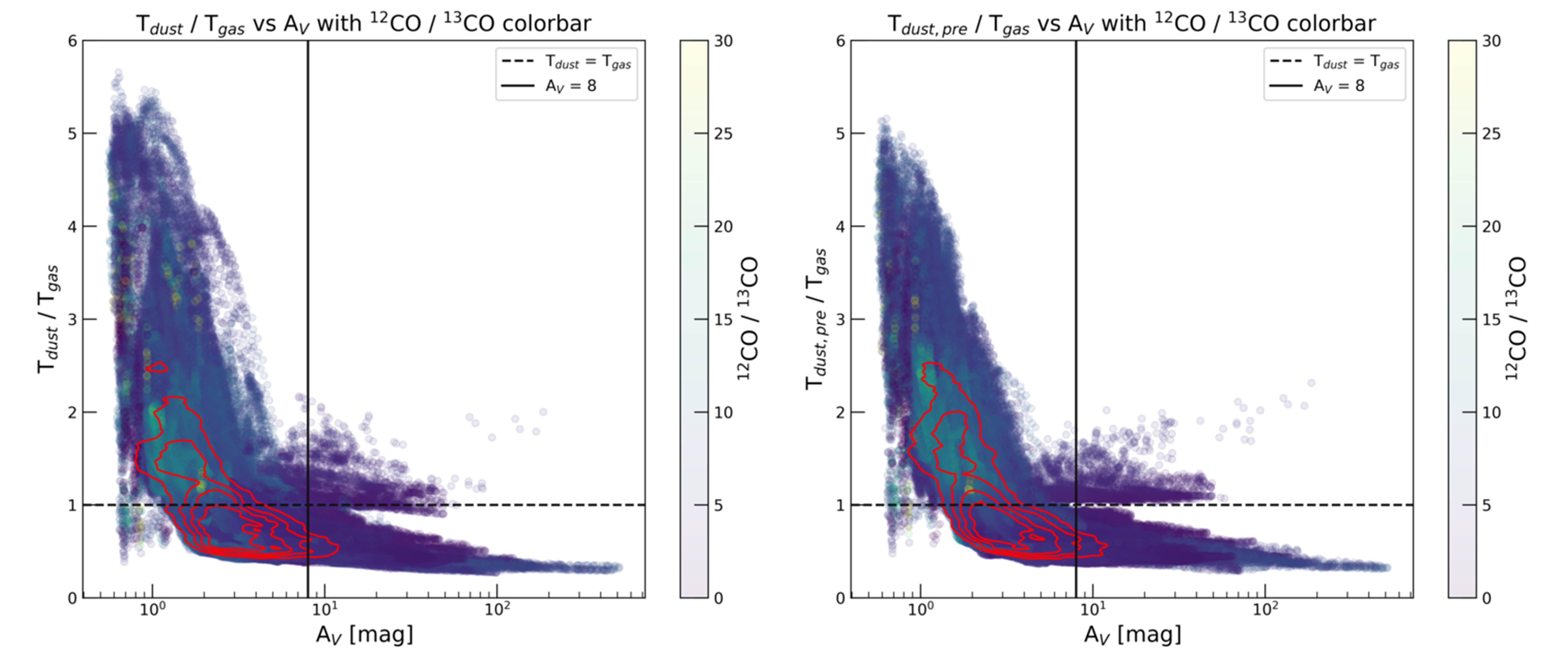}
    \caption{ Correlation plots between the ratio of the observed (left panel) and the predicted (right) dust temperature to the gas temperature and the visual extinction $A_V$ in the entire region of Orion A. Dust temperature prediction adopts the fiducial training area.
    Color scales represent the $^{12}$CO / $^{13}$CO ratio, and the red contours show the density distribution of the data points based on a kernel density estimation (KDE).
    Horizontal dashed lines show the temperature ratio of unity, while the vertical solid lines show $A_V$ = 8, which approximately separates the region between weak and strong gas-dust coupling.}
    \label{fig7}
\end{figure*}

\begin{figure*}[htbp]
    \includegraphics[width=9cm]{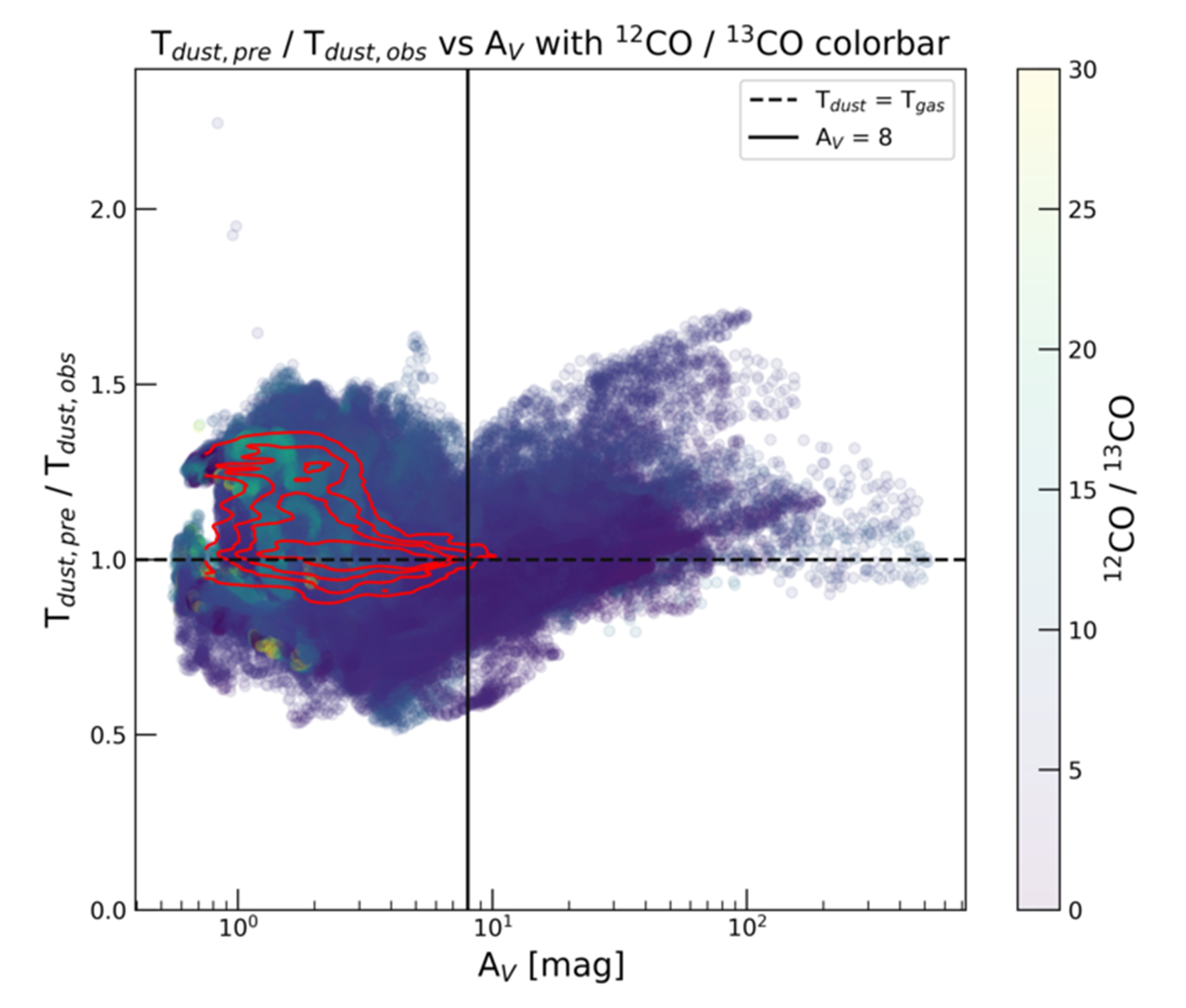}
    \caption{Correlation plot between the ratio of the predicted to the observed dust temperatures, T$_{\rm dust, pre}$ / T$_{\rm dust, obs}$, and $A_{V}$ in the entire region of Orion A. The color scale indicates the $^{12}$CO / $^{13}$CO ratio, and the red contours show the density distribution of the data points based on a kernel density estimation (KDE). Horizontal dashed lines show the temperature ratio of unity, while the vertical solid lines show $A_V$ = 8.}
    \label{fig8}
\end{figure*}

Our methodology uses molecular line data to predict dust temperatures via a machine learning approach. It is therefore natural to assume that efficient thermal coupling between gas and dust is a prerequisite for the applicability of our method. Gas and dust temperatures are expected to be well coupled in dense regions \citep{Goldsmith2001}. Conversely, in lower-density environments, the coupling becomes weaker and temperature differences emerge \citep{Koumpia2015}, making our method potentially less reliable in such regions.
In this subsection, we discuss this issue in detail.

Figure \ref{fig7} (left) shows the ratio of the observed dust temperature ($\equiv~T_{\rm dust}$) to the gas temperature ($\equiv~T_{\rm gas}$) as a function of the $A_{V}$ in the entire mapping region of Orion A.
$A_{V}$ was derived from the Herschel dust column-density map using the formula in \citet{Bohlin1978}, while $T_{\rm gas}$ was approximated from the excitation temperature as derived from the $^{12}$CO and $^{13}$CO (1--0) isotopologue line peak intensities of the 45-m data. 
A higher $A_{V}$ corresponds to a situation deeper inside the clouds and thus a higher degree of thermal coupling between gas and dust.
We found that, at a lower $A_{V}$ ($\lesssim$ 8), there is a wide dispersion of the $T_{\rm dust} / T_{\rm gas}$ ratio, and $T_{\rm dust}$ tends to be higher than $T_{\rm gas}$. These results imply less thermal coupling between dust and gas, as expected.
In such a region, the FUV radiation selectively heats dust, raising $T_{\rm dust}$ only. 
At a higher $A_{V}$ ($\gtrsim$ 8), the $T_{\rm dust} / T_{\rm gas}$ ratio converges and becomes closer to unity ($i.e.,$ efficient thermal coupling), or even lower than unity.
The transition at $A_{V}$ $\sim$8 corresponds to the typical density at which filamentary structures are known to form \citep{Andre2014}, suggesting that strong gas–dust coupling is established in these dense regions.
The presence of regions with $T_{\rm dust}$ / $T_{\rm gas}$ $<$ 1 can be explained by the difference each tracer represents: the optically thick CO line primarily traces the warmer surface layers of molecular clouds, whereas $T_{\rm dust}$ represents a column-density-weighted average temperature that decreases toward the dense cores, where cooling is more efficient.

Figure \ref{fig7} (right) shows the same figure but for the predicted dust temperature from the molecular line data with Regressor-Region-A ($\equiv T_{\rm dust, pre}$).
The $T_{\rm dust, pre} / T_{\rm gas}$ ratio shows a very similar trend to that of $T_{\rm dust} / T_{\rm gas}$.
Furthermore, the $T_{\rm dust, pre} / T_{\rm dust}$ ratio is around unity in the range of $\sim$0.5 - 1.5 at $A_{V}$ $\lesssim$ 8 (Figure \ref{fig8}).
In this regime, the data points with the value of unity are the most concentrated (red contours in Figure \ref{fig8}).
The region delineated by the red contours spatially coincides with areas exhibiting a relatively high $^{12}$CO / $^{13}$CO ratio.
This spatial correspondence suggests that the regions significantly affected by FUV radiation are primarily located at low $A_{V}$ values \citep{Nishimura2015}.
Even though in this low $A_{V}$ region, thermal coupling between gas and dust is less efficient, the dust temperature predicted from the molecular line data does not show a clear deviation from the real dust temperature.
These results imply that $T_{\rm dust, pre}$ is not primarily controlled by the gas-dust thermal coupling. On the contrary, there appears a slight increase in the $T_{\rm dust, pre} / T_{\rm dust}$ ratio to the range of 1.0 – 1.5 at $A_{V}$ $\gtrsim$8.
Such a slight increase likely reflects the higher $T_{\rm gas}$ in the high $A_{V}$ region.

The overall agreement between $T_{\rm dust}$ and $T_{\rm dust, pre}$ suggests that the machine-learning model predicting $T_{\rm dust}$ from the molecular line data is not solely controlled by the degree of the thermal gas-dust coupling.
In other words, the machine-learning approach does not simply reproduce conventional astrophysical concepts but rather captures more global correlations inherent in the observational data. Further exploration of what specific features the model learns, and a detailed comparison with traditional astrophysical analyses, would be valuable directions for future work.

\section{SUMMARY} \label{sec: summary}

We have investigated the feasibility of a machine learning technique to predict the dust temperature in the Orion A molecular cloud. The dust temperature map derived from the SED fitting to the six-band Herschel continuum data is adopted as the target value. The nine different kinds of molecular line maps made from the CO (1--0) isotopologue observations with the Nobeyama 45-m telescope are adopted as feature data. Regression models are constructed in the common training area between the dust temperature map and the molecular line map. These regressor models are then applied to regions outside the training area to evaluate their ability to predict the dust temperature maps from the molecular line data. The main results of our investigation are summarized below.

\begin{enumerate}
   \item Regressor-Region-A, our fiducial area for the regressor construction, provides a reasonable accuracy of the dust temperature predictions with the average predicted dust temperature of 20.87 K, in contrast to the average observed dust temperature of 20.25 K. This area encompasses the PDR region of the Orion Nebula and a wide temperature coverage from 14.04 K to 51.05 K. On the other hand, this area does not include temperatures lower than 14.04 K, which results in inaccurate temperature predictions in low-temperature regions outside the training area.
   
   \item The prediction accuracy of machine learning as a function of the number of pixels in the training is investigated. Regression models were constructed using various fractions of the total 337,495 mapping pixels, ranging from 0.05 \% to 50 \%. We found that even 0.05 \% of the total pixels is sufficient to provide a reasonably accurate prediction of the dust temperature map.  Furthermore, above the pixel fraction of 5 \% the R2 and average predicted dust temperatures both converge. 
   
   \item Dependence of the prediction accuracy on the selection of the training area is also studied. In addition to the fiducial area, five training areas with different dust temperature ranges and environmental conditions were selected, and separate regressors were constructed for each. Those six training areas produce different predictions of the dust temperature maps. We found that the dust temperature coverage of the training areas severely affects the prediction accuracy and that the predicted values of the dust temperatures are restricted to the original dust temperature regions of that training area. The presence or absence of PDRs in the training areas also affects the prediction accuracy for the entire regions that include such PDRs. In the regressors trained with PDR-containing regions, the $^{12}$CO / $^{13}$CO isotopic intensity ratio emerges as the most important feature, which may be related to the selective photodissociation of CO isotopologues.

   \item The present machine-learning model is capable of capturing the observed degree of thermal coupling between gas and dust.
   At a low $A_V$ region ($\lesssim$ 8), the ratio of observed dust to gas temperatures is higher than unity, suggesting selective heating of dust by the FUV radiation and less efficient gas-dust thermal coupling.
   At a higher $A_V$ ($\gtrsim$ 8), the dust-to-gas temperature ratio converges to two branches; one close to unity and the other
   even lower than unity. The brach of unity suggests
   efficient gas and dust thermal coupling, while the other branch
   likely reflects the difference of the optical depth and thus
   the layers traced by
   the molecular-line and dust-continuum emission.
   All of these trends are well reproduced by the dust temperatures
   predicted by the machine-learning model.
   These results suggest that the machine-learning prediction of the dust temperature from the molecular line data does not require a priori assumptions about the thermal coupling between gas and dust.
\end{enumerate} 

\begin{acknowledgments}
We thank the referees for their useful suggestions that improved the clarity of the paper.  
This research has made use of data from the Herschel Gould Belt Survey (HGBS) project.
This work is based in part on archival data from the Nobeyama 45-m radio telescope operated by the Nobeyama Radio Observatory (NRO), a branch of the National Astronomical Observatory of Japan.
Part of the data and data analysis were carried out using services provided by the Japanese Virtual Observatory and the Multi-wavelength Data Analysis System operated by the Astronomy Data Center, NAOJ.
K.S. is supported by NAOJ ALMA Scientific Research grant No. 2022-20A.
M.K. is supported by JSPS research grant No. JP24KJ1834.
S.T. is supported by JSPS KAKENHI grant Nos.
JP21H04495 and JP24K00674, and
by NAOJ ALMA Scientific Research grant No. 2022-20A.
This work was supported by JSPS KAKENHI Grant Number JP20K04035. This work was also supported by grants from the Mitsubishi Foundation and the Yamada Science Foundation.
\end{acknowledgments}

\bibliography{references}{}
\bibliographystyle{aasjournal}

\appendix

\section{Correlation between the dust temperature and various features}
\phantomsection \label{sec:appendixA}
\renewcommand{\thefigure}{A\arabic{figure}} 
\setcounter{figure}{0} 

\begin{figure*}[htbp]
    \includegraphics[width=18cm]{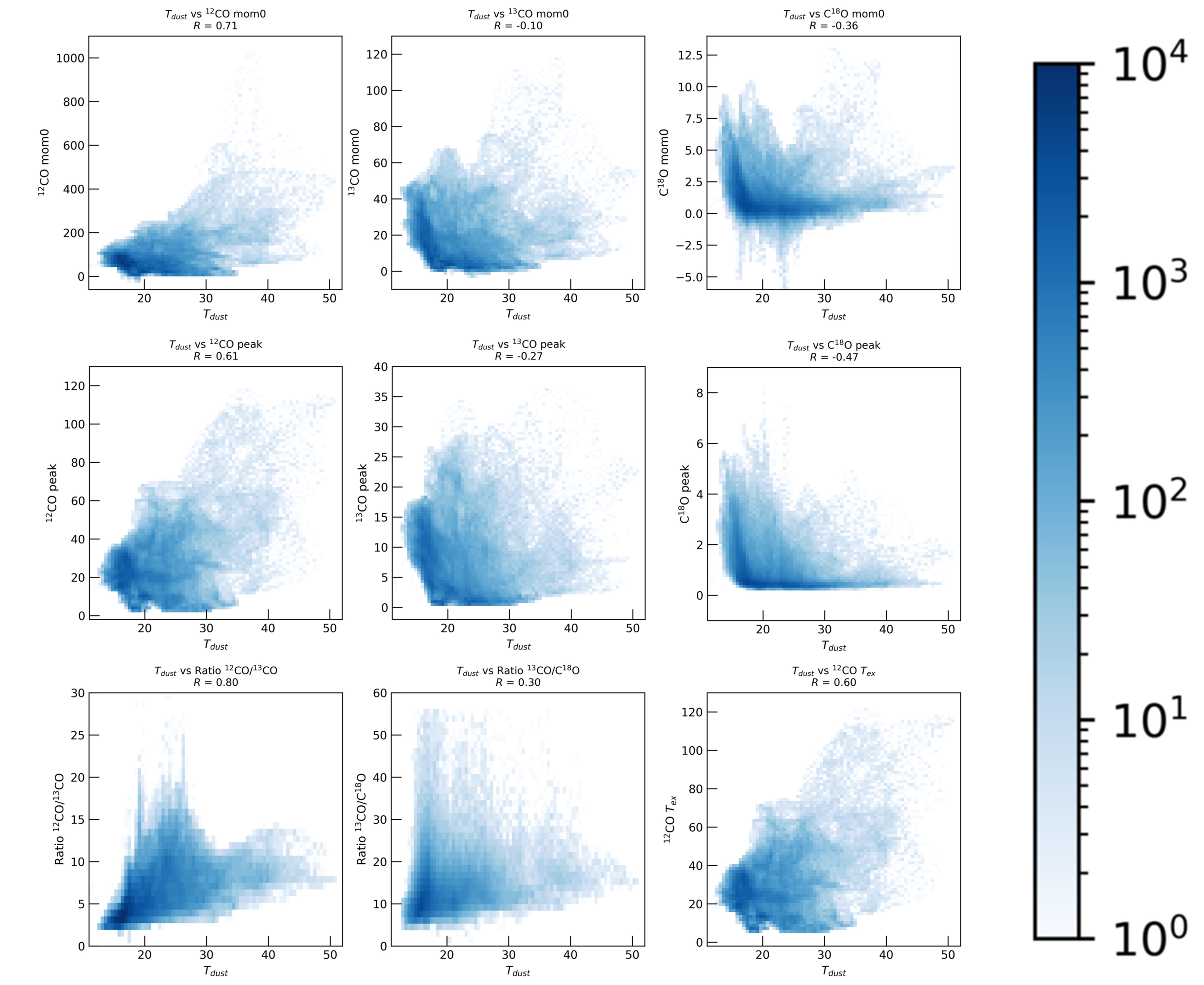}
    \caption{ 2-dimensional histograms between T$_{\rm dust}$ and the nine molecular-line features in the entire area. The blue color bar shows the number of data points.}
    \label{figA}
\end{figure*}

Figure \ref{figA} shows correlation plots between the dust temperature and
all nine molecular-line features. The relevant correlation coefficient
is also shown in each panel. Both the $^{12}$CO peak and $^{12}$CO excitation temperature exhibit visually similar distributions with respect to dust temperature,
because the $^{12}$CO excitation temperature is calculated from the
$^{12}$CO peak intensity. The machine learning regards these as
duplicated features.
Note that features that show strong correlations with the dust temperatures
(e.g., $^{12}$CO / $^{13}$CO, $^{12}$CO mom0, and $^{12}$CO peak) are also identified as
the most important features in the fiducial regression model.

\clearpage

\section{Dust Temperature Predictions with Different Fractions of the
Training Pixels}
\phantomsection \label{sec:appendixB}
\counterwithin{figure}{section}
\renewcommand{\thefigure}{B\arabic{figure}} 
\setcounter{figure}{0} 

\begin{figure}[htbp]
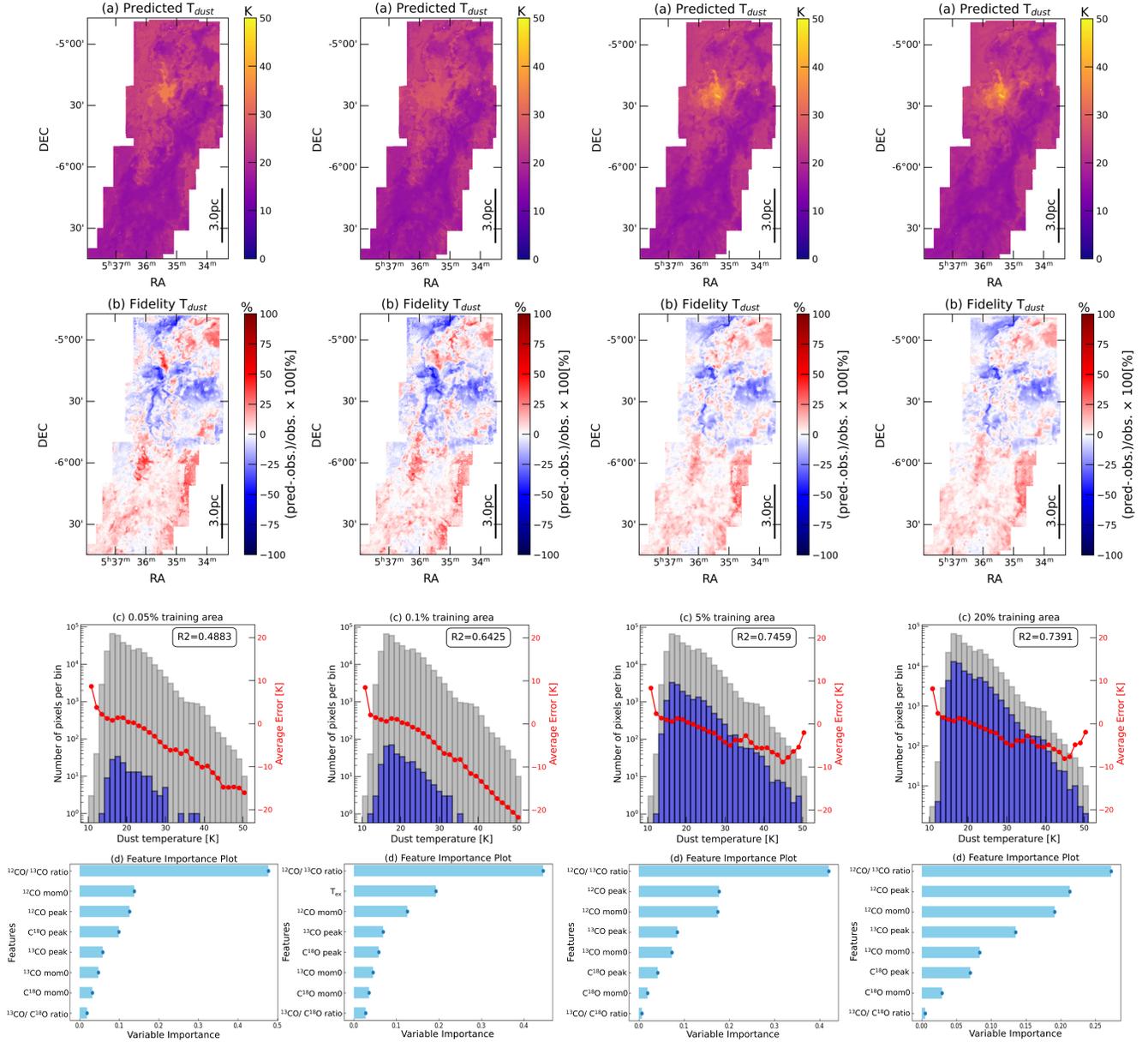

    \centering
    \includegraphics[width=1.0\textwidth]{figB1.jpg}
    \vspace{1mm} 
    \includegraphics[width=1.0\textwidth]{figB2.jpg}
    \caption{ Summary of the results of the dust-temperature prediction for four different fractions (0.05\%, 0.1\%, 5\%, and 20\%) of the training pixels. Each column shows the results of the different fractions. Each row shows (a) the predicted dust temperature maps, (b) the fidelity maps, (c) the histograms of dust temperatures in each training area (blue) with respect to that in the entire region (gray), and (d) plots of the feature importance. Red points in the third row show the average prediction error in each temperature bin.}
    \label{figB}
\end{figure}

Figure \ref{figB} shows the results of the dust temperature prediction
for four different fractions (0.05\%, 0.1\%, 5\%, and 20\%)
of the training pixels with respect to the entire pixels.

\clearpage
\section{Dust Temperature Predictions with Different Training Areas}
\phantomsection \label{sec:appendixC}
\renewcommand{\thefigure}{C\arabic{figure}}
\setcounter{figure}{0}

\begin{figure}[htbp]
    \includegraphics[width=18cm]{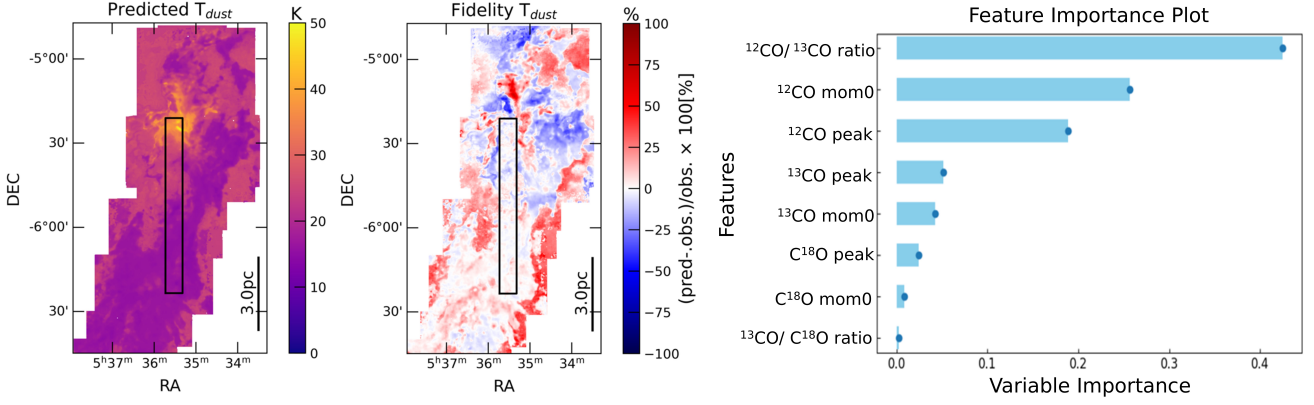}
    \caption{Maps of the predicted dust temperature (left) and fidelity (middle), and plot of the importance of the features obtained from Regressor-Region-A.} 
    \label{figC1}
\end{figure}

\begin{figure}[htbp]
    \includegraphics[width=18cm]{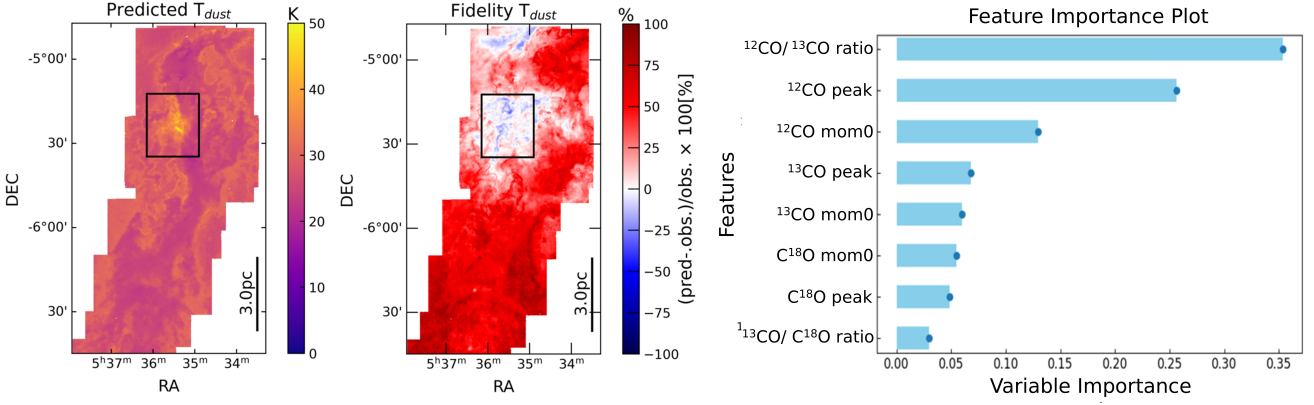}
    \caption{Same as Figure \ref{figC1} but for Regressor-Region-B.}
    \label{figC2}
\end{figure}

\begin{figure*}[htbp]
    \includegraphics[width=18cm]{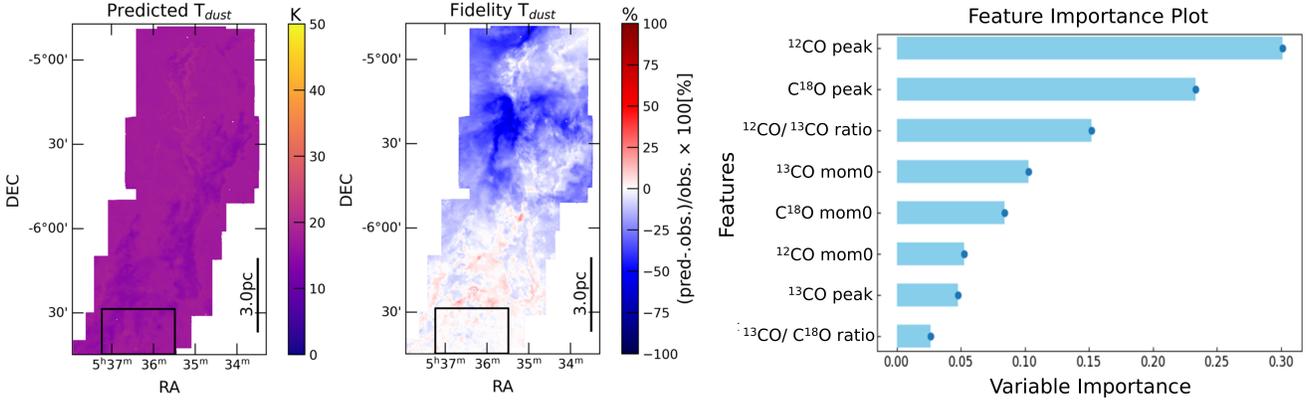}
    \caption{Same as Figure \ref{figC1} but for Regressor-Region-C.}
    \label{figC3}
\end{figure*}

\begin{figure*}[htbp]
    \includegraphics[width=18cm]{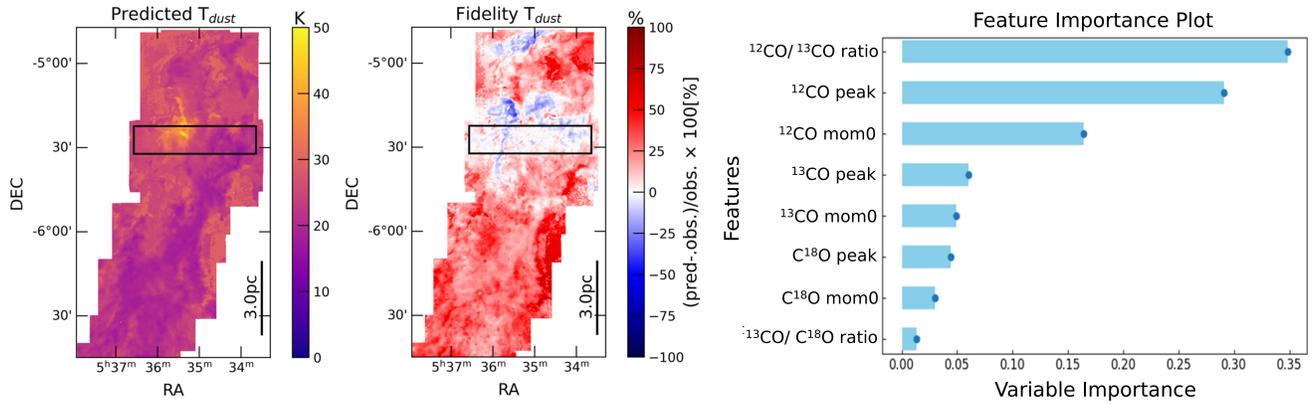}
    \caption{Same as Figure \ref{figC1} but for Regressor-Region-D.}
    \label{figC4}
\end{figure*}

\begin{figure*}[htbp]
    \includegraphics[width=18cm]{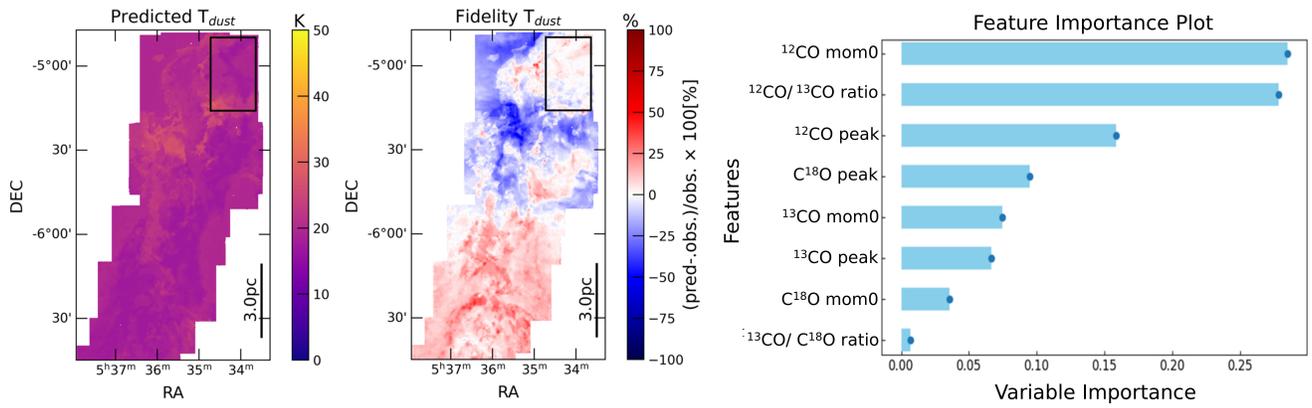}
    \caption{Same as Figure \ref{figC1} but for Regressor-Region-E.}
    \label{figC5}
\end{figure*}

\begin{figure*}[htbp]
    \includegraphics[width=18cm]{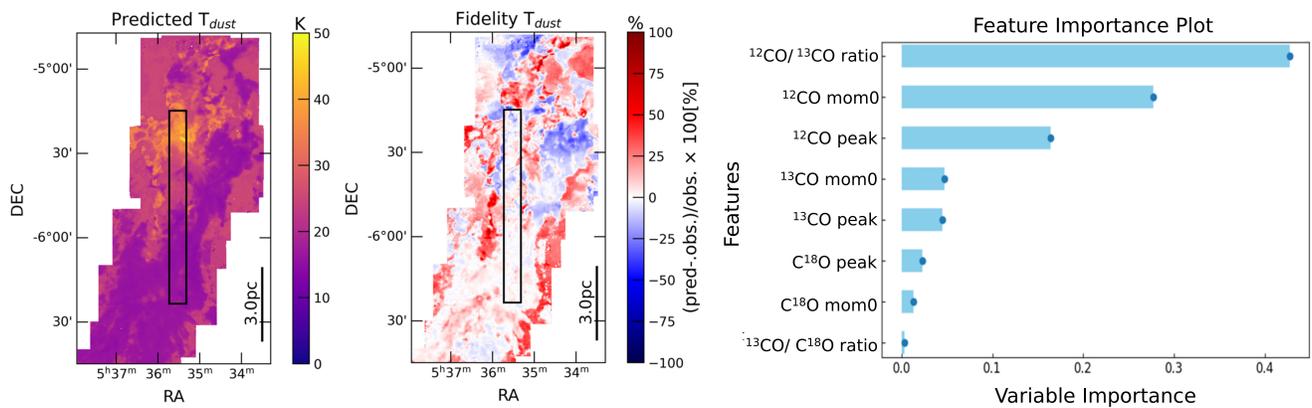}
    \caption{Same as Figure \ref{figC1} but for Regressor-Region-F.}
    \label{figC6}
\end{figure*}

Figures \ref{figC1} - \ref{figC6} show maps of the dust temperature
and fidelity predicted by each training area, and the feature importance plot.

\end{document}